\documentstyle[12pt]{article}
\topmargin - 0.8cm

\renewcommand{\theequation}{\arabic{section}.\arabic{equation}}

\catcode`@=11
\@addtoreset{equation}{section}
\catcode`@=12

\begin{document}
\title{\vskip-1.7cm \bf Wavefunction of a Black Hole and the
Dynamical Origin
of Entropy}
\author{A.O.Barvinsky${}^{*} {}^{1,2}$,\ V.P.Frolov${}^{\dag}
{}^{1,3,4}$\ and
A.I.Zelnikov${}^{\dagger} {}^{1,4}$ }
\date{}
\maketitle
\hspace{-8mm}$^{1}${\em
Theoretical Physics Institute, Department of Physics, \ University of
Alberta, Edmonton, Canada T6G 2J1}
\\ $^{2}${\em Nuclear Safety Institute, Russian Academy of Sciences ,
Bolshaya
Tulskaya 52, Moscow
113191, Russia}
\\ $^{3}${\em CIAR Cosmology Program}
\\ $^{4}${\em P.N.Lebedev Physics Institute, Russian Academy of
Sciences ,
Leninskii Prospect 53, Moscow
117924, Russia}
\begin{abstract}
Recently \cite{FrNo:93a,FrNo:93b} it was proposed to explain the
dynamical
origin of the entropy of a black hole by identifying  its dynamical
degrees of
freedom   with states of quantum fields propagating in the
black-hole's
interior. The present paper contains the further development of this
approach.
The no-boundary proposal (analogous to the Hartle-Hawking no-boundary
proposal
in quantum cosmology) is put forward for defining the wave function
of a black
hole. This  wave function is a functional on the configuration space
of
physical fields (including the gravitational one) on the
three-dimensional
space with the Einstein-Rosen bridge topology.  It is shown that in
the limit
of small perturbations on the Kruskal background geometry the
no-boundary wave
function coincides with the Hartle-Hawking vacuum state. The
invariant
definition of inside and outside modes is proposed. The density
matrix
describing the internal state of a black hole is obtained by
averaging over the
outside modes. This density matrix is used to define the entropy of a
black
hole, which is to be divergent. It is argued that the quantum
fluctuations of
the horizon which are internally present in the proposed formalism
may give the
necessary cut-off and provide a black hole with the finite entropy.
\end{abstract}
\\
$^{*}$Electronic address: barvi@phys.ualberta.ca\\
$^{\dag}$Electronic address: frolov@phys.ualberta.ca\\
$^{\dagger}$Electronic address: zelnikov@phys.ualberta.ca

\newpage
\baselineskip6.8mm
\section{Introduction}
\hspace{\parindent}
According to the  thermodynamical analogy in  black hole physics,
the
entropy of a black hole  is defined as $\mbox{\boldmath $S$}^H =A^H
/(4l_{\mbox{\scriptsize{P}}}^2)$,
where   $A^H$   is   the   area   of   a   black   hole   surface
and
$l_{\mbox{\scriptsize{P}}}=m_{\mbox{\scriptsize{P}}}^{-1}=(\hbar
G/c^3~)^{1/2}$   is   the   Planck length \cite{Beke:72,Beke:73}. The
Hawking
discovery \cite{Hawk:74,Hawk:75} of the  black hole thermal radiation
confirmed
the status of thermal properties of a black hole. Four laws of black
hole
physics formulated in \cite{BaCaHa:73} show that a black hole can be
considered
as a thermodynamical system and the entropy  of a black hole plays
essentially
the same role as the entropy in the 'usual' physics, e.g., it shows
up to
which extent the  energy contained in a black hole  can be used to
produce
work.   The generalized  second law  (i.e. the statement that the sum
$\mbox{\boldmath $S$}=\mbox{\boldmath $S$}^H +\mbox{\boldmath $S$}^m$
of a
black hole entropy and  the entropy $\mbox{\boldmath $S$}^m$ of the
outside
matter cannot decrease) implies that in the case when a black hole is
part of a
thermodynamical system the thermodynamical laws will be
self-consistent only if
 the black hole  entropy is considered on equal footing with the
entropy of the
'usual' matter \cite{Beke:72,Beke:73,Beke:74} (see also
\cite{ThPrMa:86,NoFr:89,Wald:92,FrPa:93} and references therein).
Gedanken
experiment proposed by  York \cite{York:83} in which a black hole is
located
inside a heated cavity gives a nice example showing that such
parameters of a
black hole as  a heat capacity and entropy have a well defined
physical
meaning.

The formal derivation of thermal properties of a black hole is
usually
performed in the framework of the Euclidean approach initiated by
Gibbons and
Hawking \cite{GiHa:76,Hawk:79}. It  implies the existence of the
thermodynamical ensemble of black holes characterized by the
canonical
partition function at finite temperature $T=1/\beta$
	\begin{equation}
	Z(\beta)={\rm Tr}\,e^{-\beta\hat H},    \label{1.1}
	\end{equation}
where $\hat H$ is the Hamiltonian of the full gravitational system.
The known
functional representation of finite temperature field theory in terms
of the
Euclidean quantum theory, directly extrapolated to quantum gravity,
allows one
to rewrite (\ref{1.1}) as a Euclidean path integral over 4-geometries
and
matter fields. The evaluation of this integral by the steepest
descent method
determines  $Z(\beta)$ and, in particular, gives $T=1/8\pi M$. A
refined
version of this approach which emphasizes the role of boundary
conditions was
developed in \cite{York:86,BrBrWhYo:90,BrYo:93a,BrYo:93b}. In the
framework of
this approach  the construction of the microcanonical partition
function within
the Lorentzian context was analysed and  a revised issue of stability
for the
gravitational ensemble at a given temperature and given boundary
quasilocal
characteristics was given.

Although the Euclidean approach allows to obtain the correct value
for the
black hole entropy, it does not elucidate the number of questions.
Mainly this
is a question of the origin of the thermodynamical partition
(\ref{1.1}) which
is assumed to be given for granted. In other words, it does not
specify the
physical degrees of freedom inaccessible for observation for an
external
observer, their tracing out in the pure quantum state of the whole
gravitational system leading to the loss of information, emergence of
entropy
and the density matrix corresponding to (\ref{1.1}).  In particular,
the
conventional Euclidean approach to gravitational thermodynamics
simply does not
leave room for a black hole interior, for it is located completely
outside of
the real Euclidean section of the complex Schwarzschild geometry.

Despite some promising attempts
\cite{Beke:73,ThPrMa:86,York:83,Beke:80,ZuTh:85,Hoof:85}, the
dynamical
(statistical mechanical) origin of a black hole entropy has not been
well understood.  According to  the `standard' interpretation,  the
entropy of
a black hole   is  considered  as  a  logarithm of the number of
distinct
ways    that    the   hole    might have   been     made
\cite{ThPrMa:86,ZuTh:85}. This interpretation  is not satisfactory.
Soon after
the black hole formation neither external nor internal observer can
see or
affect these states and hence it does not make sense to interpret
them as usual
dynamical degrees of freedom which specify the state of the system at
the
chosen moment of time. The problem of relation of the black hole
entropy to the
loss of information about the initial state of a collapsing body is a
part of
very important problem of information loss in the black hole
evaporation
\cite{Hawk:76a}. We shall not consider the problem of information
loss in the
present paper and restrict ourselves to the problem of dynamical
origin of a
black hole entropy.

York \cite{York:83} proposed to identify the dynamical degrees of
freedom of a
black hole with its quasi-normal modes. But the entropy of the
quasinormal
modes excited at a given moment of time is much smaller $
\mbox{\boldmath
$S$}^H =A^H  /(4l_{\mbox{\scriptsize{P}}}^2)  $.
In order to obtain the required large number 't Hooft \cite{Hoof:85}
proposed a
"brick wall model". In this approach the entropy of a black hole is
identified
with the entropy of a thermal gas located outside a black hole and
supported in
equilibrium by a heated wall located at small distance outside the
horizon. The
value of the gap parameter in this model is chosen by equating the
entropy of
the gas outside the wall to the entropy of a black hole. The relation
of the
"brick wall" model to the results obtained from the first principles
remains
unclear.

Recently \cite{FrNo:93a,FrNo:93b}  a new approach to the problem of
black hole
entropy was proposed. According to this approach the dynamical
degrees  of  freedom  of  a  black  hole are identified with those
modes of
physical fields which are located inside the horizon and  which
cannot be seen
by a distant observer.   It
was   shown   that   the main  contribution to  the entropy is given
by
thermally excited  `invisible' modes propagating in the close
vicinity of  the
horizon. The so defined one-loop entropy of a black hole is formally
divergent
and requires a cut-off \cite{fn1}.
This divergence is caused by a sharp boundary of the invisible region
and it
arises already in the similar flat spacetime calculations
\cite{BoKoLeSo:86,Sred:93}.  The natural cut-off arises because of
the quantum
fluctuations of the horizon.  A   calculation  based  on  a simple
estimate of
the horizon fluctuations \cite{FrNo:93a,FrNo:93b} yields a value of
the
entropy which is   in   good  agreement    with the   usually
adopted    value
$A^H/(4l^2_{\mbox{\scriptsize{P}}})$ .

There are two important problems which naturally arise in connection
with these
results. 1) How to generalize the calculation of the entropy in order
to
include the quantum fluctuations of the horizon in a self-consistent
way? 2)
How to combine the developed approach with the calculations of the
black hole
entropy based on the Euclidean space approach?

It looks like that it is impossible to solve these problems without
developing
the quantum scheme which includes the quantization of the
gravitational field.
In this paper we present an approach which might be regarded as an
attempt to
fill the gaps in the theoretical foundation  of black-hole
thermodynamics. It
consists of i) the proposal for the pure quantum state of the black
hole, ii)
the invariant dynamical criterion for the separation of its quantum
degrees of
freedom into observable ones and those inaccessible for an exterior
observer
and iii) averaging over the latter variables which leads to the
density matrix
of a black hole and the dynamical origin of its entropy. We also
briefly
discuss the recently proposed idea \cite{SuUg:94} that the problem of
the
entropy is related to the problem of renormalization of the
gravitational
constant.

\section{Dynamical Degrees of Freedom of a Black Hole}
\hspace{\parindent}
The object we are interested in is a black hole which arises as a
result of the
gravitational collapse. For simplicity we assume that a black hole is
non-rotating and spherically symmetric.  Denote by $\mbox{\boldmath
$\Sigma$}_0$ a spacelike or null global Cauchy surface and denote by
$\partial
B$ the intersection of a surface $\mbox{\boldmath $\Sigma$}_0$ with
the event
horizon $H^+$ of the black hole. The state of our system (a black
hole and
fields in its vicinity) can be characterized by giving the values of
gravitational and other fields on a chosen surface $\mbox{\boldmath
$\Sigma$}_0$. It is evident that  the states of the gravitational and
other
fields  located inside $\partial B$ have no influence on the further
evolution
of the black hole exterior. For states of particles and fields which
fall into
the a black hole from exterior region the energy $E$ defined by means
of the
timelike Killing vector $\xi$ is always positive. (For particles
$E\equiv
-\xi^{\mu}p_{\mu}$, where $p^{\mu}$ is its momentum.) Besides these
states
inside the black hole there exist states with negative total energy
$E<0$. Such
states, located inside the black hole at $\mbox{\boldmath $\Sigma$}$,
will be
considered as its internal degrees of freedom \cite{fn3}.

The study of internal degrees of freedom of a black hole is
complicated because
 a  surface $\mbox{\boldmath $\Sigma$}$  crosses the singularity.
There exist
more convenient approach which greatly simplifies the consideration.
A lone
black hole at late time (i.e. long after a black hole formation)  is
almost
stationary, i.e., its state can be described as the classical static
(Kruskal)
metric and small perturbations (fields excitation) propagating on
this
background. Analytical continuations of a static black hole solution
defines
maximally extended solution which is known as eternal black hole
metric.  If
$\mbox{\boldmath $\Sigma$}_0$
is chosen at late time one can also trace back in time all the fields
excitations present in the vicinity of  $\mbox{\boldmath $\Sigma$}_0$
so that
the problem of specifying the states of a black hole can be
reformulated as a
problem for an eternal black hole. Technically the latter is much
simpler, so
that we use this approach.

The  Kruskal metric for eternal black hole reads
	\begin{equation}
	ds^2 =- \frac{32M^3}{r}\exp \left[-(\frac{r}{2M}-1)\right]
       dU dV +r^2 d\Omega ^2 , \label{2.1}
	\end{equation}
	\begin{equation}
	UV =(1-\frac{r}{2M})\exp (\frac{r}{2M}-1) \ .
\label{2.2}
	\end{equation}
Denote by  $\mbox{\boldmath $\Sigma$}$  a global Cauchy surface
defined by the
equation $U+V=0$. It has wormhole topology $\mbox{\boldmath
$R$}\times
\mbox{\boldmath $S$}^2$.  This is a well-known Einstein-Rosen bridge
connecting
two asymptotically flat three-dimensional spaces. The discrete
isometry
$U\rightarrow  -U, V\rightarrow -V$ transforms the surface
$\mbox{\boldmath
$\Sigma$}$ onto itself, so that one asymptotically flat region
($\mbox{say
\boldmath $\Sigma$}_+$) is mapped onto another ($\mbox{say \boldmath
$\Sigma$}_-$). Localized states with $E<0$ being traced back in time
in the
Kruskal geometry cross $\mbox{\boldmath $\Sigma$}_-$, while states
with $E>0$
cross $\mbox{\boldmath $\Sigma$}_+$.

A remarkable property of the Kruskal-Schwarzschild metric (\ref{2.2})
is that
it can be considered as a real Lorentzian-signature section of the
complex
manifold parametrized by the real radial $r,\,0\leq r<\infty,$ and
complex time
$z$ coordinates:
	\begin{eqnarray}
	&&z=\tau+it,\\
	&&U=-\left(\frac r{2M}-1\right)^{1/2}
	{\rm exp}\left\{\frac12 \Big(\frac r{2M}-1\Big)
	+i\,\frac{z-2\pi M}{4M}\right\},\\
	&&V=\ \ \left(\frac r{2M}-1\right)^{1/2}
	{\rm exp}\left\{\frac12 \Big(\frac r{2M}-1\Big)
	-i\,\frac{z-2\pi M}{4M}\right\}.
	\end{eqnarray}
Sectors $R_+$ and $R_-$ of the Kruskal metric are generated by the
following
segments in the complex plane of $z$
	\begin{eqnarray}
	&&R_+: z=2\pi M+it,\,-\infty<t<\infty,\nonumber \\
	&&R_-: z=-2\pi M+it,\,-\infty<t<\infty,
	\end{eqnarray}
and analytically joined by the real Euclidean section $E$
	\begin{eqnarray}
	&&E: z=\tau,\,-2\pi M\leq \tau\leq 2\pi M.    \label{2.7}
	\end{eqnarray}
Here $t$ is a usual time-like Killing coordinate in the Schwarzschild
metric,
while $\tau$ is its Euclidean analogue playing the role of the
angular
coordinate in the Gibbons-Hawking black hole instanton periodic with
the period
$\beta=8\pi M$:
	\begin{eqnarray}
	ds^2_E=\left(1-\frac{2M}r\right)\,d\tau^2
	+\left(1-\frac{2M}r\right)^{-1}dr^2
	+r^2d\Omega^2.                             \label{2.8}
	\end{eqnarray}
The Euclidean section (\ref{2.7}) represents a half-period part of
this
instanton, the boundary $\mbox{\boldmath $\Sigma$}_{+}\cup
\mbox{\boldmath
$\Sigma$}_{-}$ of which at $\tau_{\pm}=\pm 2\pi M$ represents the
Einstein-Rosen bridge of the above type. At this boundary the
Euclidean section
analytically matches with the Lorentzian sectors $R_+$ and $R_-$ on
the Penrose
diagram of the Kruskal metric.

To determine the propagation of small perturbations on the background
of an
eternal black hole one must specify initial data at $\mbox{\boldmath
$\Sigma$}_0$. It is evident that the data at  $\mbox{\boldmath
$\Sigma$}_-$
(the part of  $\mbox{\boldmath $\Sigma$}_0$ lying inside the black
hole) do not
influence the black hole exterior. That is why these data can be
identified
with internal degrees of freedom of a black hole. The gravitational
perturbations can be dealt with in the same manner. Our main idea can
be
described as follows. Fix a three-dimensional manifold with a
wormhole topology
$\mbox{\boldmath $R$}\times \mbox{\boldmath $S$}^2$ and consider any
three-dimensional metrics on it which posses two asymptotically flat
regions.
Consider also configuration of matter fields on this manifold. This
space of
3-geometries and matter fields will be considered as a configuration
space for
our problem. We introduce a wave function of a black hole as a
functional on
this configuration space. It should be stressed that the metric and
fields at
the 'internal' part $\mbox{\boldmath $\Sigma$}_-$ of space are to be
considered
as defining the internal state of a black hole and hence they will be
identified with its internal degrees of freedom.

Our proposal for the quantum state of a black hole is a "no-boundary"
wavefunction of 3-geometry and matter fields on such a surface
$\mbox{\boldmath
$\Sigma$}=\mbox{\boldmath $R$}\times \mbox{\boldmath $S$}^2$ given by
the
Euclidean path integral of Hartle and Hawking over 4-geometries and
spacetime
matter-field configurations bounded by $\Sigma$ and 4-dimensional
asymptotically flat and empty infinity.

Obviously, the above picture is only an illustration of the general
method we
shall propose here. In the full quantum gravity incorporating the
coupling of
matter with the gravitational field (what is usually called a
self-consistent
back reaction of quantized matter on semiclassical background) many
features of
the Schwarzschild solution do not persist. There is no Killing
symmetries, the
very notion of the bifurcation surface of the Killing horizon
separating
physical variables into observable and unobservable ones does not
exist and
should be dynamically determined on the ground of some invariant
criterion. In
this paper we propose such a criterion which allows one to formulate
the notion
of the black hole horizon subject to quantum vibration ( the horizon
{\it
zitterbewegung} ) and to calculate  its quantum dispersion. The
latter quantity
is very important in gravitational thermodynamics \cite{York:83}, for
it,
apparently, provides a self-consistent high energy cutoff for the
one-loop
entropy \cite{FrNo:93a,FrNo:93b}.

It should be emphasized that the quantum state of the black hole we
advocate
here is merely a {\it proposal}, and we must verify its validity by
comparing
its consequences with the known properties of the conventional
gravitational
thermodynamics. For this purpose we first show that, semiclassically,
this
state generates the black-hole Hartle-Hawking vacuum \cite{HaHo} for
the
particle excitations of all spins (including graviton) and produces
by the
procedure of the above type the thermal density matrix with the
temperature
$T=1/8\pi M$. We then estimate the dominant contribution to the
one-loop
entropy of the black hole, which proves to be divergent in the
vicinity of the
horizon . This means that one-loop effects must necessarily be
included in the
consideration of the self-consistent gravitational thermodynamics.

\section{No-boundary wavefunction of a black hole}
\hspace{\parindent}
The no-boundary wavefunction was first proposed by Hartle and Hawking
\cite{HH,H} in the context of quantum cosmology as a path integral
	\begin{eqnarray}
	\mbox{\boldmath $\Psi$}(^3\!g(\mbox{\boldmath $x$}),
	\varphi(\mbox{\boldmath $x$}))=
	\int D \,{}^4\!g \ D\mbox{\boldmath $\phi$}
	{\rm e}^{\!\!\phantom{0}^
	{\textstyle -\mbox{\boldmath $I$}[\,{}^4\!g ,\mbox{\boldmath
$\phi$}\,]}}
	                    \label{3.1}
	\end{eqnarray}
of the exponentiated gravitational action $\mbox{\boldmath
$I$}[\,{}^4\!g
,\mbox{\boldmath $\phi$}\,]$ over Euclidean 4-geometries and
matter-field
configurations on a compact spacetime $\mbox{\boldmath $M$}$ with a
boundary
$\partial\mbox{\boldmath $M$}$. The integration variables are subject
to the
conditions
	$(^3\!g(\mbox{\boldmath $x$}),\,
	\varphi(\mbox{\boldmath $x$})),\;\;
	\mbox{\boldmath $x$}
	\in\partial\mbox{\boldmath $M$}$, -- the collection of
3-geometry and boundary
matter fields  on $\partial\mbox{\boldmath $M$}$, which are just the
argument
of the wavefunction (\ref{3.1}).

This construction was also applied in the asymptotically-flat case
\cite{Hartle-as} when $\mbox{\boldmath $M$}$ represents a noncompact
4-dimensional half-space whose boundary consists of two components,
$\partial\mbox{\boldmath $M$}=\mbox{\boldmath
$R^3$}\bigcup\partial\mbox{\boldmath $M$}_{\infty}$: infinite
3-dimensional
hyperplane $\mbox{\boldmath $R^3$}$ carrying the field argument of
the
wavefunction  and the 4-dimensional asymptotically-flat and empty
infinity
$\partial\mbox{\boldmath $M$}_{\infty}$. The latter is a singular
component of
the spacetime boundary and its boundary conditions are in certain
sense trivial
and do not enter the argument of the wavefunction.

We propose the quantum state of a black hole which is a modification
of this
asymptotically-flat no-boundary wavefunction of Hartle. It is given
by
eq.(\ref{3.1}) where the total boundary
	\begin{eqnarray}
	\partial\mbox{\boldmath $M$}=
	\mbox{\boldmath $\Sigma$}
	\cup\partial\mbox{\boldmath $M$}_{\infty}  \label{3.3}
	\end{eqnarray}
has instead of the hyperplane above the hypersurface with the
topology of the
Einstein-Rosen bridge
	\begin{eqnarray}
	\mbox{\boldmath $\Sigma$}=
	\mbox{\boldmath $R$}\times
	\mbox{\boldmath $S$}^2                   \label{3.4}
	\end{eqnarray}
connecting two asymptotically flat 3-dimensional spaces.

The construction (\ref{3.1}) - (\ref{3.4}) forms a topological part
of the
definition for the no-boundary wavefunction. Apart from that the
expression
(\ref{3.1}) signifies nothing unless we specify the meaning of the
integration
measure
$\,{}^4\!g ,\mbox{\boldmath $\phi$}\,$ . We also need to determine
the physical
inner product with respect to which one can calculate  the
expectation values
and matrix elements for a given wavefunction. In the context of the
Lorentzian
spacetime the problems have a solution which is based on the
quantization of
true physical variables \cite{ADM,Kuh,BPon} and can be constructively
realized
at least within the semiclassical loop expansion \cite{B:GenSem}.
This
quantization leads to the standard Faddeev-Popov integration measure
\cite{Faddeev} in the functional integral (\ref{3.1}) and to its
analogue in
the physical inner product for the wavefunction
$\mbox{\boldmath $\Psi$}(^3\!g(\mbox{\boldmath $x$}),\,
\varphi(\mbox{\boldmath $x$}))$ in the representation of local
spatial 3-metric
tensor and matter fields. The measure in this physical inner product
is
nontrivial.
It is roughly the Faddeev-Popov measure in the configuration space of
fields
taken on a single spatial surface of the spacetime. The measure
incorporates
the gauge fixing procedure and effectively restricts the integration
to the
subset of true configuration-space coordinates among the dynamically
redundant
set $(^3\!g(\mbox{\boldmath $x$}),\,
\varphi(\mbox{\boldmath $x$}))$ \cite{BPon,B:GenSem}:
	\begin{eqnarray}
	^3\!g(\mbox{\boldmath $x$}),\,
	\varphi(\mbox{\boldmath $x$})\rightarrow
	\varphi=(\,g^{T}(\mbox{\boldmath $x$}),\,
	\varphi(\mbox{\boldmath $x$})\,).          \label{3.5}
	\end{eqnarray}
The geometrical content of the local gravitational variables can be
very
different depending on the choice of gauge and generally represents
certain two
dynamically independent degrees of freedom $g^{T}(\mbox{\boldmath
$x$})$  per
spatial point. They originate from solving the gravitational
constraints and
imposed gauge conditions for the original gravitational phase-space
variables
$^3\!g(\mbox{\boldmath $x$}),\,^3\!p(\mbox{\boldmath $x$})$ in terms
of
$g^{T}(\mbox{\boldmath $x$})$ and physical conjugated momenta
$p_{T}(\mbox{\boldmath $x$})$
\footnote
{Complicated gauge conditions can generally mix the original
gravitational
variables with matter ones, but here we disregard this possibility
and consider
only the case of (\ref{3.5}) when the gravitational physical degrees
of freedom
are disentangled from the gravitational sector of the theory. Still,
in view of
this fact, we use the neutral symbol $\varphi$ to denote the full set
of
physical configuration coordinates without emphasizing their metric
or matter
content.}.

The wavefunction can be constructed directly in the representation of
physical
variables (\ref{3.5}), $\Psi(\varphi)$. In this representation the
physical
inner product has a trivial form
	 \begin{equation}
	 <\!\Psi_{1}\,|\,\Psi_{2}\!>=\int d\varphi\,
	 \Psi_{1}^{*}\,(\varphi)\,\Psi_{2}(\varphi),     \label{3.6}
	 \end{equation}
that provides the unitary dynamics of $\Psi(\varphi)=\Psi(\varphi,t)$
with the
physical Hamiltonian whose functional form arises from the ADM
reduction
(\ref{3.5})
\footnote
{The unitary map between the Dirac-Wheeler-DeWitt wavefunctions
$\mbox{\boldmath $\Psi$}(^3\!g(\mbox{\boldmath
$x$}),\,\varphi(\mbox{\boldmath
$x$}))$ and wavefunctions of true physical variables
$\Psi(\varphi,t)$ is
discussed in much detail in \cite{B:GenSem} both at the level of the
path
integral and operatorial quantizations.}.
For this reason, we shall formulate our proposal (\ref{3.1}) -
(\ref{3.4}) for
the wavefunction of a black hole in the representation of physical
variables
\footnote
{In the cosmological context the no-boundary wavefunction in such a
representation was considered in \cite{Hartle-Sch} and also
constructed as a
unifying link between the Lorentzian and Euclidean quantum gravity
theories in
\cite{tunnelI}.}.  In this representation the wave function of a
black hole is
given by the path integral of the form (\ref{3.1}), but with the
physical
configuration coordinates (\ref{3.5}) fixed at
$\partial\mbox{\boldmath $M$}$
instead of the 3-metric components of the dynamically redundant set
$(^3\!g(\mbox{\boldmath $x$}),\,
\varphi(\mbox{\boldmath $x$}))$
	\begin{eqnarray}
	\Psi(\varphi)=\int \limits_
   {  {\textstyle\,
	\phi  |_{\Sigma}
	=\varphi}  }
         D\phi\,
	\,{\rm e}^{\!\!\phantom{0}^{\textstyle -\mbox{\boldmath
$I$}[\,\phi\,]
	}} \ .                                   \label{3.7}
	\end{eqnarray}
Here the integration goes over those spacetime histories of physical
ADM fields
$\phi=\phi(x)$ that generate the Euclidean 4-geometries
asymptotically flat at
the infinity $\partial\mbox{\boldmath $M$}_{\infty}$ of spacetime
and match
$\varphi$ on its "dynamically active" boundary (\ref{3.4}).
$\mbox{\boldmath
$I$}[\,\phi\,]$ is the Lagrangian gravitational action in terms of
these
fields. The integration measure $D\phi$ involves the local functional
measure
\cite{tunnelI} the structure of which is not very important for our
purposes.

As it was mentioned above, the nature of physical degrees of freedom
depends on
the choice of gauge in the ADM reduction procedure. To effectively
operate with
the physical wavefunction, we have to fix this gauge and  perform the
reduction
(\ref{3.5}). Here we use a York gauge \cite{York} which consists of
the
condition
	\begin{eqnarray}
	{\rm tr}\,^3\!p(\mbox{\boldmath $x$})\equiv
	\,g^{ab}(\mbox{\boldmath $x$})\,
	p_{ab}(\mbox{\boldmath $x$})=0,                  \label{3.8}
	\end{eqnarray}
selecting a spacetime foliation by minimal surfaces (of vanishing
mean
extrinsic curvature ${\rm tr}\,K(\mbox{\boldmath $x$})=0$), and some
other
three conditions fixing the coordinatization of these surfaces. A
distinguished
nature of this gauge consists in the fact that, in contrast to a
majority of
other gauges, it does not suffer from the problem of Gribov copies
invalidating
the physical reduction when the latter is considered globally in
phase space of
the theory
\footnote
{This property actually poses a dilemma of York gauges versus the
third
quantization of gravity, a strong motivation for the latter being
rooted in the
problem of Gribov copies problem in quantum gravity theory (see a
discussion in
\cite{B:GenSem}).
}.
This property of the York gauge follows from a strong theorem of
\cite{York-Murch} on the uniqueness of a solution of the Lichnerowicz
equation
for the conformal factor in the conformal decomposition of a 3-metric
\cite{York}, provided positive-energy condition holds for matter
fields.

As known \cite{York,Isenberg-Marsden}, the physical degrees of
freedom in the
York gauge can be represented by the two variables characterizing the
conformally-invariant part of the 3-metric $tilde
g_{ab}(\mbox{\boldmath $x$})$
(in some gauge fixing of the 3-dimensional spatial diffeomorphisms)
and the
conjugated transverse traceless momenta $tilde p^{ab}(\mbox{\boldmath
$x$})$,
while the conformal mode $\Phi(\mbox{\boldmath $x$})$ of the full
3-metric
	\begin{eqnarray}
	 g_{ab}(\mbox{\boldmath $x$})=
	\Phi^4(\mbox{\boldmath $x$})\;
	\tilde g_{ab}(\mbox{\boldmath $x$})  \label{3.9}
	\end{eqnarray}
follows from the solution of the Lichnerowicz equation which is just
the
Hamiltonian gravitational constraint rewritten in the conformal
decomposition
of the above type
	\begin{eqnarray}
	&&(\tilde\Delta-\frac18\,^3\!\tilde R)\,
	\Phi+\frac18\,C\Phi^{-7}+2\pi\,
	\tilde T^*_*\,\Phi^{-3}=0,       \label{3.10}\\
	&&C\equiv \,
	\tilde p^{ab}\,\tilde p_{ab} / \tilde g .    \label{3.11}
	\end{eqnarray}
Here $\tilde T^*_{\star}=\Phi^{-8}T^*_*$ is a conformally rescaled
energy --
Hamiltonian density -- of matter fields and tilde denotes the
quantities
calculated in the conformal metric $\tilde g_{ab}$
\footnote{
In the geometrically invariant language, the physical content of
$\tilde
g_{ab}$ can be described by the conformally invariant
transverse-traceless
tensor of York $\beta^{ab}$ \cite{York}.
}.
In the linearized approximation the physical gravitational variables
in the
York gauge are the transverse-traceless part of the linear
excitations $h_{a
b}$ and their conjugated transverse-traceless momenta
\footnote
{We assume that, without loosing the generality, the spatial gauge
conditions
fixing the coordinatization of metrically perturbed $\mbox{\boldmath
$\Sigma$}$
can be chosen as transversality of $h_{ab}$. The variables
$(h_{ab}^T,\,p_{T}^{ab})$ are conformally related to their tilded
conformally-invariant analogues the transversality and tracelessness
of which
holds with respect to $\tilde g_{ab}$.
}.
	\begin{eqnarray}
	 (g^{T},\,p_{T})=(h_{ab}^T,\,p_{T}^{ab}), \label{3.12}
	 \end{eqnarray}
In  a semiclassical approximation the wave function of a black hole
	\begin{eqnarray}
	\Psi(\varphi)=P\,{\rm e}^{\!\!\phantom{0}^
	{\textstyle -\mbox{\boldmath $I$}[\,\phi (\varphi)\,]}}
\label{3.13}
	\end{eqnarray}
is dominated by the classical action at the extremal of equations of
motion
$\phi (\varphi)$ subject to boundary conditions $\varphi$ on
$\mbox{\boldmath
$\Sigma$}$. It also includes the preexponential factor $P$
accumulating the
result of integration over quantum field deviations from the
extremal. The
physical variables $\varphi$ given by eqs.(\ref{3.5}) and
(\ref{3.12}) are
treated by perturbations and the Euclidean action $\mbox{\boldmath
$I$}[\,\phi
(\varphi)\,]$ is to be expanded in powers of $\varphi$. To obtain the
lowest-order term $\mbox{\boldmath $I$}[\,\phi (0)\,]$, notice that
the
boundary 3-geometry on $\mbox{\boldmath $\Sigma$}$ (\ref{3.9}) has,
in virtue
of (\ref{3.10}) a conformal factor satisfying the homogeneous
conformally-invariant equation in three dimensions. As shown in
Appendix A, it
gives for asymptotically flat boundary conditions exactly the
spherically
symmetric metric of the Einstein-Rosen bridge, characterized by a
single
constant -- the mass $M$ of the black hole. The extremal of the
Euclidean
vacuum Einstein equations $\phi (0)$ satisfies asymptotically flat
boundary
conditions at $\partial\mbox{\boldmath $M$}_{\infty}$. The
corresponding
solution is just one half of the Schwarzschild gravitational
instanton of mass
$M$ with the four-dimensional metric (\ref{2.8}) for $-2\pi
M\leq\tau\leq 2\pi
M$.  The classical  action on this half of instanton reduces to the
contribution of the surface term at $\partial \mbox{\boldmath
$M$}_{\infty}$ of
the classical Einstein gravitational action
	\begin{eqnarray}
	\mbox{\boldmath $I$}[\,\phi (0)\,] = \frac1{8\pi}
	\int_{{\textstyle\partial} \mbox{\boldmath $M$}_{\infty}}
	K\sqrt{h}d^3 x
	= 2\pi M^2.                 \label{3.13a}
	\end{eqnarray}

The expansion of $\mbox{\boldmath $I$}[\,\phi (\varphi)\,]$ in powers
of
$\varphi$ on the background of $\phi (0)$ shows that the linear-order
term
vanishes due to the equations of motion for the background and the
vanishing of
the extrinsic curvature of $\mbox{\boldmath $\Sigma$}$ (the latter
property
guarantees the absence of the surface terms). Therefore the leading
contribution to the semiclassical wavefunction (\ref{3.13}) takes the
form
	\begin{eqnarray}
	\Psi(\varphi,M)=P\,{\rm e}^{\!\!\phantom{0}^{
	{\textstyle - 2\pi M^2\! - \mbox{\boldmath $I$}_2[\,\phi
(\varphi)\,]}}},
     \label{3.14}
	\end{eqnarray}
where $\mbox{\boldmath $I$}_2[\,\phi (\varphi)\,]$ is a quadratic
term of the
action in the linearized physical fields (\ref{3.5}) and
(\ref{3.12}).

Thus, our no-boundary wavefunction of a black hole turns out to be a
functional
of the local gravitational and matter degrees of freedom
$\varphi({\mbox{\boldmath $x$}})$, parametrized by a global variable
-- the
gravitational mass of the Einstein-Rosen bridge $M$. Obviously, if we
include
$M$ into the configuration space of the black hole, the dependence of
the
wavefunction on it will describe the probability distribution of
black holes
with different masses in this quantum state.  A naive inclusion of
$M$ into the
ADM phase space of the theory in the York gauge does not seem to be
fully
consistent. However, it was recently performed in more general
context  by
K.Kuchar \cite{Kuchar} who persuasively advocated that $M$ has a
conjugated
momentum $P_M$, so that $(M,P_M)$ can be a subject to standard
canonical
quantization and incorporate as their quantum state an {\it
arbitrary} function
of the black-hole mass $M$. Thus, the proposed $M$-dependent
no-boundary
wavefunction can be regarded as a first example of such a quantum
state of a
black hole (or, more precisely, of the quantum Einstein-Rosen bridge)
\footnote
{The variables $(M,P_M)$ of Ref.\cite{Kuchar} have the nature of
angle-action
variables in their canonical action. The variable $M$ plays the role
of the
positive conserved energy and the "angle" $P_M$ linearly grows in
time with the
speed determined by the way the observer anchors the spacetime
foliation at
spatial infinity with his physical clock. This means that, strictly
speaking,
their quantum state can not be absolutely arbitrary function of $M$
and, in
particular, it  can generate the discrete spectrum of masses for the
quantum
Einstein-Rosen bridge. \cite{Beke:74,Mukhanov,Berezin}.
}.
In what follows, however,  we shall consider $M$ as an external
parameter not
entering the argument of the wavefunction and, correspondingly,
excluded from
the phase space and the Hilbert space of the theory.  Therefore, up
to
$M$-dependent normalization, the semiclassical wavefunction of the
black hole
will be dominated by its ${\rm exp}\,(-\mbox{\boldmath $I$}_2[\,\phi
(\varphi)\,])$ part, describing the dynamics of local degrees of
freedom. In
the next section we show that it represents their Hartle-Hawking
vacuum on the
background of the Kruskal-Schwarzschild geometry.

\section{Hartle-Hawking vacuum state}
\hspace{\parindent}
We demonstrate now the calculation of (\ref{3.14}) and its vacuum
properties on
a simple example of a scalar field
$\phi\,(x)=\phi\,(\tau,{\mbox{\boldmath
$x$}})$ with the quadratic action
	\begin{eqnarray}
	\frac12 \int d^4
x\,g^{1/2}\Big(g^{\mu\nu}\partial_{\mu}\phi\,
	\partial_{\nu}\,\phi+\xi\,R\,\phi^2\Big).
\label{4.1}
	\end{eqnarray}
The generalization to fields of higher spins in the quadratic
approximation is
obvious. This action generates on the Euclidean section (\ref{2.7})
with the
metric (\ref{2.8}) the linear equations of motion
	\begin{eqnarray}
	\left\{-g^{1/2}g^{\tau\tau}\frac{d^2}{d\tau^2}-
	\partial_ag^{1/2}g^{ab}\partial_b\right\}
	\phi\,(\tau,{\mbox{\boldmath $x$}})=
        0,\;\;{\mbox{\boldmath $x$}}=x^a,\;\;a=1,2,3, \label{4.2}
	\end{eqnarray}
which must be solved subject boundary conditions
$\varphi=\varphi\,({\mbox{\boldmath $x$}})$ on its boundary
$\mbox{\boldmath
$\Sigma$}$ to give the extremal $\phi_*(\varphi)$ of eq.(\ref{3.14}).
On the
Schwarzschild background with $R=0$ the nonminimal interaction does
not
contribute to the equations. In what follows we denote the boundary
fields on
the two asymptotically flat parts of the Einstein-Rosen bridge
$\mbox{\boldmath
$\Sigma$}_{\pm}$ by $\varphi_{\pm}$
	\begin{eqnarray}
	\phi\,(x)\,\Big|_{\textstyle\,
	\mbox{\boldmath $\Sigma$}_{\pm}}
	\equiv\phi\,(\pm \beta/4,{\mbox{\boldmath $x$}})
	=\varphi_{\pm}({\mbox{\boldmath $x$}}).
\label{4.3}
	\end{eqnarray}
With this notation the solution to (\ref{4.2}) can be written as a
decomposition
	\begin{eqnarray}
	\phi_*(\tau,{\mbox{\boldmath
        $x$}})=\sum_{\lambda}\Big\{\,\varphi_{\lambda,+}
	u_{\lambda,-}(\tau,{\mbox{\boldmath $x$}})+
	\varphi_{\lambda,-}
	u_{\lambda,+}(\tau,{\mbox{\boldmath $x$}})\,\Big\}
\label{4.4}
	\end{eqnarray}
in the basis functions of this equation
	\begin{eqnarray}
	&&u_{\lambda,\pm}(\tau,{\mbox{\boldmath $x$}})=
	\frac{{\rm sinh}\,(\beta/4\mp\tau)\,\omega}
	{{\rm sinh}\,(\beta/2)}\,R_{\omega l m A}({\mbox{\boldmath
$x$}}),\;\;
	\lambda=(\omega,l,m,A),                        \label{4.5}
	\end{eqnarray}
containing the set of spatial harmonics $R_{\omega l m
A}({\mbox{\boldmath
$x$}})$ -- eigenfunctions of the following eigenvalue problem
	\begin{eqnarray}
	\partial_a\Big(g^{1/2}g^{ab}\partial_b
	R_{\omega l m A}({\mbox{\boldmath $x$}})\,\Big)=
	-g^{\tau\tau}g^{1/2}\omega^2
	R_{\omega l m A}({\mbox{\boldmath $x$}})
\label{4.6}
	\end{eqnarray}
originating from the separation of variables in (\ref{4.2}). These
eigenfunctions are enumerated by a set of continuous $\omega>0$ and
discrete
$(l,m,A)$ labels, among which $l$ and $m$ are the usual quantum
numbers of
spherical harmonics and the label $A=1,2$ is responsible for two
possible
directions of propagation along the radial coordinate. As shown in
Appendix B,
these spatial harmonics can be chosen real. They are required to be
regular at
the horizon $r=2M$ and the spatial infinity, have a positive definite
spectrum
$\omega^2>0$ and satisfy the orthonormality and completeness
conditions
	\begin{eqnarray}
	&&\int d^3 x\,g^{\tau\tau}g^{1/2}
R_{\lambda}({\mbox{\boldmath $x$}})\,
	R_{\lambda'}({\mbox{\boldmath
$x$}})=\delta_{\lambda\lambda'},\\
	&&\sum_{\lambda} R_{\lambda}({\mbox{\boldmath $x$}})\,
	R_{\lambda}({\mbox{\boldmath $x$}}')=
	\frac{\delta({\mbox{\boldmath
$x-x'$}})}{g^{\tau\tau}g^{1/2}}.
	\end{eqnarray}
Here, as in (\ref{4.5}), we use a condensed notation $\lambda$ for
the full
collection of quantum numbers, the summation over which implies the
following
measure
	\begin{eqnarray}
	\sum_{\lambda}(...)\equiv
	\int_{0}^{\infty}d\omega\sum_{l,m,A}(...),\;\;
	\delta_{\lambda\lambda'}\equiv
	\delta(\omega-\omega')\,\delta_{ll'}\delta_{mm'}\delta_{AA'}
	\end{eqnarray}

In view of these relations the coefficients $\varphi_{\lambda,\pm}$
in
(\ref{4.4}) are just the decomposition coefficients of the fields
(\ref{4.3})
in the basis of spatial harmonics
	\begin{eqnarray}
	\varphi_{\pm}({\mbox{\boldmath $x$}})=
	\sum_{\lambda} \varphi_{\lambda,\pm}
R_{\lambda}({\mbox{\boldmath $x$}}).
	\end{eqnarray}

Substituting (\ref{4.4}) into (\ref{4.1}), integrating by parts with
respect to
Euclidean time and taking into account the equations of motion
(\ref{4.2}), one
finds that the Euclidean action reduces to the following quadratic
form in
$\varphi_{\lambda,\pm}$ (cf. a similar derivation in
\cite{Laflamme}):
	\begin{eqnarray}
	\mbox{\boldmath $I$}_2(\varphi_{+},\varphi_{-})=
	\frac12\,\sum_{\lambda}\left
	\{\,\frac{\omega_{\lambda}\,{\rm
cosh}(\beta\,\omega_{\lambda}/2)}
	{{\rm sinh}(\beta\,\omega_{\lambda}/2)}
	\,(\varphi_{\lambda,+}^2+\varphi_{\lambda,-}^2)
	-\frac{2\,\omega_{\lambda}}{{\rm sinh}(\beta\,
	\omega_{\lambda}/2)}
	\,\varphi_{\lambda,+}\varphi_{\lambda,-}\right\} \label{4.11}
	\end{eqnarray}
that can be diagonalized by the following reparametrization to new
variables
$f_{\lambda,\pm}$
	\begin{eqnarray}
	&&\!\!\!\!\!\!\!\!\!\!\!\!\!\!\!\!\!\!\!\!\!\!\!\!\varphi_{\pm
}=
\frac{f_{+}\pm f_{-}}{\sqrt 2},\label{4.12}\\
&&\!\!\!\!\!\!\!\!\!\!\!\!\!\!\!\!\!\!\!\!\!\!\!\!\mbox{\boldmath $I$}_2(\varphi_{+},\varphi_{-})=
\bar{\mbox{\boldmath $I$}}_2(f_{+},f_{-})=
\frac12\,\sum_{\lambda}\omega_{\lambda}\left
\{\,{\rm tanh}(\beta\,\omega_{\lambda}/4)\,f_{\lambda,+}^2
+\frac1{{\rm
tanh}(\beta\,\omega_{\lambda}/4)}\,f_{\lambda,-}^2
	\!\right\}.
	\end{eqnarray}

The wavefunction (\ref{3.14}) rewritten in the new representation
(\ref{4.12})
is a gaussian state which is obviously a vacuum
	\begin{eqnarray}
	&&\Psi\,(\varphi_{+},\varphi_{-})=\bar\Psi(f_{+},f_{-})=
	P\,{\rm e}^{\!\!\phantom{0}^
	{\textstyle -\mbox{\boldmath $\overline{I}$}_2
(f_{+},f_{-})}},
\label{4.14}\\
	&&\tilde a_{\pm}\,\bar\Psi\,(f_{+},f_{-})=0,
	\end{eqnarray}
of the following creation-annihilation operators (we omit for brevity
the label
$\lambda$ in the definition of $\tilde a_{\pm}$ below as well as in
$\omega=\omega_{\lambda}$):
	\begin{eqnarray}
	&&\tilde a_{+}=\frac 1{\sqrt 2}\,
	\left[\;\Big(\omega\,{\rm
tanh}\frac{\beta\omega}4\Big)^{\!-1/2}
	\frac \partial{\partial f_{+}}\,+\,
	\Big(\omega\,{\rm tanh}\frac{\beta\omega}4\Big)^{1/2}
	f_{+}\,\right],
\label{4.16}    \\
	&&\tilde a_{+}^\dagger=\frac 1{\sqrt 2}\,
	\left[-\Big(\omega\,{\rm
tanh}\frac{\beta\omega}4\Big)^{\!-1/2}
	\frac \partial{\partial f_{+}}+
	\Big(\omega\,{\rm tanh}\frac{\beta\omega}4\Big)^{1/2}
	f_{+}\,\right],   \nonumber\\
	&&\tilde a_{-}=\frac 1{\sqrt 2}\,
	\left[\,\Big(\,\frac 1\omega\,{\rm
tanh}\frac{\beta\omega}4\Big)^{1/2}
	\frac \partial{\partial f_{-}}+
	\Big(\,\frac 1\omega\,{\rm
tanh}\frac{\beta\omega}4\Big)^{\!-1/2}
	f_{-}\right],
\label{4.17}              \\
	&&\tilde a_{-}^\dagger=\frac 1{\sqrt 2}\,
	\left[-\Big(\,\frac 1\omega\,{\rm
tanh}\frac{\beta\omega}4\Big)^{1/2}
	\!\frac \partial{\partial f_{-}}+\!
	\Big(\,\frac 1\omega\,{\rm
tanh}\frac{\beta\omega}4\Big)^{\!-1/2}
	\!f_{-}\right],  \nonumber
	\end{eqnarray}
subject to standard commutation relations
	\begin{eqnarray}
	\left[\tilde a_{\lambda,\pm},
	\tilde
a_{\lambda'\pm}^\dagger\right]=\delta_{\lambda\lambda'}
	\end{eqnarray}
(all the other commutators are vanishing). For our purposes another
choice of
creation-annihilation operators is more useful, differing from
(\ref{4.16})-(\ref{4.17}) by the linear transformation not mixing the
positive
and negative frequencies
	\begin{eqnarray}
	a_{\lambda,\pm}=\frac{\tilde a_{\lambda,+}
	\pm \tilde a_{\lambda,-}}{\sqrt 2},\;\;\;
	a_{\lambda,\pm}\bar\Psi\,(f_+,f_-)=0.         \label{4.21}
	\end{eqnarray}

To give a particle interpretation for the obtained vacuum state we
must
construct the propagating physical modes corresponding to
$a_{\lambda,\pm}$.
For this purpose consider the $\mbox{\boldmath $\Sigma$}_{\pm}$ parts
of
$\mbox{\boldmath $\Sigma$}$ as the initial Cauchy surfaces in the
right $(R_+)$
and left $(R_-)$ wedges of the Lorentzian Kruskal-Schwarzschild
spacetime. In
these two causally disconnected regions lying to the future of
$\mbox{\boldmath
$\Sigma$}_{\pm}$ one can construct two scalar field theories with the
Lagrangians -- the Lorentzian versions of (\ref{4.1})
	\begin{eqnarray}
	L_{\pm}=\int_{\mbox{\boldmath $\Sigma$}_{\pm}} d^3 x\,
	{\cal L}\,(\phi,\partial\phi)=
	\frac 12\,\sum_{\lambda}\,(\dot\varphi_{\lambda,\pm}^2
	-\omega_{\lambda}^2 \varphi_{\lambda,\pm}^2),
	\end{eqnarray}
which take such a form provided the corresponding spacetime fields
evolving
correspondingly in $R_+$ and $R_-$ are decomposed in spatial
harmonics with
time-dependent coefficients $\varphi_{\lambda,\pm}(t)$,
$\dot\varphi_{\lambda,\pm}\equiv d\varphi_{\lambda,\pm}(t)/dt$. At
the quantum
level, in the coordinate representation of $\varphi_{\lambda,\pm}$
the
creation-annihilation operators $b_{\lambda,\pm}$ of these two
theories
associated with positive-negative frequency decomposition in the
Killing time
$t$ look as follows
	\begin{eqnarray}
	\sqrt 2\,b_{\lambda,\pm}=\frac 1{\sqrt\omega}\,
	\frac\partial{\partial\varphi_{\lambda,\pm}}
	+\sqrt\omega\,\varphi_{\lambda,\pm},          \label{4.23}\\
	\sqrt 2\,b_{\lambda,\pm}^\dagger=-\frac 1{\sqrt\omega}\,
	\frac\partial{\partial\varphi_{\lambda,\pm}}
	+\sqrt\omega\,\varphi_{\lambda,\pm}            \label{4.24}
	\end{eqnarray}
and correspond to the following choice of positive-frequency basis
functions
	\begin{eqnarray}
	&&w_{\lambda,+}(x)\;\Big|_{R_+}=
	{\rm e}^{\textstyle -i\omega_\lambda t}
	R_{\lambda}({\mbox{\boldmath $x$}}),\;\;\;
	w_{\lambda,+}(x)\;\Big|_{R_-}=0      \label{4.25}\\
	&&w_{\lambda,-}(x)\;\Big|_{R_-}=
	{\rm e}^{\textstyle\, i\omega_\lambda t}
	R_{\lambda}({\mbox{\boldmath $x$}}),\;\;\;
	w_{\lambda,-}(x)\;\Big|_{R_+}=0         \label{4.26}
	\end{eqnarray}
(one should remember that the Schwarzschild Killing time coordinate
is past
pointing in $R_-$ and $w_{\pm}$ by construction have zero initial
data on
$\mbox{\boldmath $\Sigma$}_{\mp}$).

This is a matter of a simple algebra, using the reparametrization
(\ref{4.12}),
to show that the operators (\ref{4.23}) are related to (\ref{4.21})
by a
nontrivial Bogolyubov transformation which mixes the positive and
negative
frequencies
	\begin{eqnarray}
	b_{\pm}=\Big(2\,{\rm sinh}\frac{\beta\omega}2\Big)^{-1/2}
	\left[\,{\rm e}^{\textstyle\,\beta\omega/4}a_{\pm}+
	{\rm e}^{\textstyle\,-\beta\omega/4}a_{\mp}^\dagger\right]
	\end{eqnarray}
and generates, in terms of $w_{\pm}$, the basis functions
$v_{\lambda,\pm}(x)$
associated with the creation annihilation operators $a_{\lambda,\pm}$
of our
vacuum quantum state (\ref{4.21})
	\begin{eqnarray}
	v_{\pm}=\Big(2\,{\rm sinh}\frac{\beta\omega}2\Big)^{-1/2}
	\left[\,{\rm e}^{\textstyle\,\beta\omega/4}w_{\pm}+
	{\rm e}^{\textstyle\,
	-\beta\omega/4}w_{\mp}^*\right].       \label{4.28}
	\end{eqnarray}
This is a well-known transformation relating the Killing vacua,
$(b_{\lambda,\pm},\;w_{\lambda,\pm}(x))$, in the right ($R_+$) and
left ($R_-$)
wedges of the Kruskal diagram to the Hartle-Hawking vacuum,
$(a_{\lambda,\pm},\;v_{\lambda,\pm}(x))$, of quantum fields on the
maximally
extended black hole spacetime \cite{Hartle-Hvac}
\footnote
{The doubled set of field modes in Schwarzschild-Kruskal spacetime
and their
thermofield nature \cite{Umezawa}  was noticed by W.Israel
\cite{Israel}, this
observation being further developped within the context of the
Euclidean path
integral in \cite{Laflamme} (see also \cite{Frolov-Mart}).
}.
The latter is defined by the condition that its basis functions
$v_{\lambda,\pm}(x)$ contain only positive frequencies with respect
to affine
parameter on both horizons of the black hole metric. This property
follows from
eqs.(\ref{4.25})-(\ref{4.26}), (\ref{4.28}) and the asymptotic
behaviors of
$w_{\lambda,\pm}(x))$ at the horizon (see Appendix B)
        \begin{eqnarray}
	&&w_{+}(x)\;\Big|_{{\,}R_+}=\left\{\begin{array}{l}
	C^{\rm past} (-U)^{\;\textstyle 4M\omega i},
	\;\;\;\;\;x\rightarrow H_{+}^{\rm past},\\
	C^{\rm future} (V)^{\textstyle -4M\omega i},
	\;\;\;x\rightarrow H_{+}^{\rm future},
	\end{array}\right.
	\\
	&&w_{-}(x)\;\Big|_{R_-}=\left\{\begin{array}{l}
	(C^{\rm past})^*\,(U)^{\textstyle -4M\omega i},
	\;\;\;\;\;x\rightarrow H_{-}^{\rm future},\\
	(C^{\rm future})^*\,(-V)^{\textstyle 4M\omega i},
	\;\;\;x\rightarrow H_{-}^{\rm past},
	\end{array}\right.,
	\end{eqnarray}
where $C^{\rm past}$ and $C^{\rm future}$ are some complex
coefficients and
$H_{\pm}^{\rm past}$ and $H_{\pm}^{\rm future}$ are past and future
horizons of
the $\pm$ wedges of the Kruskal diagram. Substituting these behaviors
into
(\ref{4.28}) one finds
	\begin{eqnarray}
	v_{+}(x)\,\Big|_{\,\textstyle H_{+}^{\rm past}
	\cup H_{-}^{\rm future}}=
	C^{\rm past}\Big(2\,{\rm sinh}\frac{\beta\omega}2\Big)^{-1/2}
	\left[\,\theta(-U)\,{\rm e}^{\textstyle\,2\pi M\omega}
	(-U)^{\;\textstyle 4M\omega i}\right.
	\nonumber\\
	+\left.\theta(U)\,{\rm e}^{\textstyle\,-2\pi M\omega}
	U^{\;\textstyle 4M\omega i}\right]
	\end{eqnarray}
which is a basis function with the needed positive frequency behavior
matching
with the analyticity in the lower half of the complex $U$-plane
\cite{Hartle-Hvac}. The same proof holds for another horizon of the
Kruskal
diagram.

As it was mentioned above, similar considerations apply to fields of
all
possible spins. Thus, the proposed no-boundary wavefunction of a
black hole
represents the Hartle-Hawking vacuum state of linearized field
excitations of
all physical fields.

\section{One-Loop Contribution to Entropy of a Black Hole}
\hspace{\parindent}
Now we return to the problem of a black hole entropy. According to
the above
procedure of separating the physical variables into observable and
unobservable
ones, the proposed wave generates the density matrix of a black hole
interior
as a functional trace  $\mbox{Tr}_{+}$ over the values of the field
$\varphi
_{+}{{\mbox{\boldmath $(x)$}}} $ outside the horizon
	\begin{eqnarray}
	\rho (\varphi '_{-} ,\varphi _{-} )
	=\mbox{Tr}_{+}\,|\Psi><\Psi|
	\equiv \int {D\varphi_{+}}
	\,\Psi^{*}(\varphi '_{-},\varphi_{+})
         \,\Psi(\varphi_{-},\varphi_{+}) .             \label{5.1}
	\end{eqnarray}
It gives rise to the entropy of the black hole
	\begin{eqnarray}
	\mbox{\boldmath $S$}=-\mbox{Tr}\,[ \,\hat{\rho}
\ln{\hat{\rho}} \,]= - \int{D
\varphi
	\langle\varphi  ({\mbox{\boldmath $x$}})\mid
	\hat{\rho} \ln{\hat{\rho}}
        \mid  \varphi  ({\mbox{\boldmath $x$}})\rangle} .
\label{5.2}
	\end{eqnarray}
Up to normalization, the wavefunction defined by the path integral
over
physical degrees of freedom (\ref{3.7}) in the $\tau$-foliation of
the
Euclidean spacetime, $-\beta/4<\tau<\beta/4$,  actually represents
the heat
kernel or the matrix element between the configurations $\varphi_{-}$
and
$\varphi_{+}$ of the Euclidean "evolution" operator
${\rm exp}(-\beta\hat{H}/2)$
	\begin{eqnarray}
	\mbox{\boldmath $\Psi$}  (\varphi _{-} ,\varphi _{+} )
        =\exp{\left(\mbox{\boldmath $\Gamma$} \over 2\right)}<
\varphi _{-}\mid
	\exp{(-{\beta \over 2}\,\hat{H})}
	\mid \varphi _{+} >\,\Big|_{\,\beta=8\pi M},   \label{5.5}
	\end{eqnarray}
where $\hat{H}$ is a physical Hamiltonian of the system. In the full
nonperturbative treatment of the problem this Hamiltonian is a
complicated
functional of physical degrees of freedom, numerically coinciding
with the ADM
surface integral, while in the linearized approximation (relevant to
the
one-loop order of semiclassical expansion) it is just an additive sum
of
quadratic Hamiltonians of fields of all spins on the
Schwarzschild-Kruskal
background. In particular, for a scalar field, we shall consider
here, it is
the following expression generated by the Lagrangian (\ref{4.1})
	\begin{eqnarray}
	H(\pi,\varphi)={1 \over 2}\int{d^3 x
	\left[
 	{1 \over {g^{\tau\tau} g^{1 \over 2}}}\pi ^2 +g^{1 \over 2}
\left[ g^{a b}
	\partial _a \varphi \partial _b \varphi +  \xi R \phi^2
\right]
	\right]} . \label{H}
	\end{eqnarray}
Using the composition law, $\int
d\varphi_{+}\,|\varphi_{+}><\varphi_{+}|=\hat
1$,  we obtain from (\ref{5.5}) the following representation for the
density
matrix
	\begin{eqnarray}
	\rho (\varphi '_{-} ,\varphi _{-} )=
         \exp{(\mbox{\boldmath $\Gamma$}) }< \varphi '_{-}\mid
	\exp{(-8\pi M\, \hat{H})} \mid \varphi _{-} >    ,
	\end{eqnarray}
where $\mbox{\boldmath $\Gamma$}$ is defined from  the normalization
conditions
	\begin{eqnarray}
	\mbox{Tr}_{-} \hat\rho=\int D\varphi_{-} \
\rho(\varphi_{-},\varphi _{-})  =
1.
	\end{eqnarray}
In addition to the density matrix $\hat{\rho}$ it is convenient also
to define
a more general object $\hat{\rho}_\beta$  which depends on the
arbitrary
parameter $\beta$ independent of the black hole mass
	\begin{eqnarray}
	&&\hat{\rho}=\hat{\rho}_\beta \,\Big| _{\,\beta=8\pi M} \ \ ,
\\
	&&\rho_\beta (\varphi '_{-} ,\varphi _{-} )=
	 < \varphi '_{-}\mid  \hat{\rho}_\beta  \mid \varphi _{-} >
	= \exp{(\mbox{\boldmath $\Gamma$}_\beta) }< \varphi '_{-}\mid
	\exp{(- \beta  \hat{H})} \mid \varphi _{-} >.   \nonumber
\label{5.11}
	\end{eqnarray}
In the one-loop approximation we have
	\begin{eqnarray}
	< \varphi '_{-}\mid\exp{(-\beta  \hat{H})} \mid \varphi _{-}
>  =
	\left[\,\det  {1 \over {2 \pi} }
	{{\partial ^2  \mbox{\boldmath $I$}(\varphi '_{-},\varphi
_{-})
	 }\over {\partial \varphi '_{-} \partial \varphi _{-}}}
\right] ^{1/2}
	 \ \exp{[-\mbox{\boldmath $I$}(\varphi '_{-},\varphi _{-})]}
,
	\end{eqnarray}
where the Euclidean Hamilton-Jacobi function $\mbox{\boldmath
$I$}(\varphi
'_{-},\varphi _{-})$ is given by the equation (\ref{4.11}) which in
the
coordinate representation of a scalar field gives rise to the
following kernel
of the Van Vleck - Morette functional matrix
	\begin{eqnarray}
	{{\partial ^2 \mbox{\boldmath $I$}(\varphi '_{-},\varphi
_{-})}\over
        {\partial \varphi '_{-}({\mbox{\boldmath $x$}})
	\partial \varphi _{-}({\mbox{\boldmath $y$}})}} =
g^{\tau\tau} g^{1 \over 2}
	{{\hat{\omega}} \over {\sinh{\beta\hat{\omega}}}}
         \delta ({\mbox{\boldmath $x$}}-{\mbox{\boldmath $y$}})
	\end{eqnarray}
with the operator of frequency $\hat\omega$ defined  on a spatial
3-dimensional
hypersurface as
	\begin{eqnarray}
	\hat{\omega}&=&\left[ -
	{1 \over {g^{\tau\tau} g^{1 \over 2}}}\partial _a g^{1 \over
2} g^{a
b}\partial _b
	+ {1 \over {g^{\tau\tau}}} \xi R
	\right]^{ 1\over 2} ; \hskip 3cm  a,b,...={1,2,3,}\\
	g &\equiv& \det g_{\mu\nu} = g_{\tau\tau} \det g_{a b} ;
	 \hskip 4.4cm \mu,\nu,...={0,1,2,3}.
	\end{eqnarray}
The normalization factor of eq.(\ref{5.11}) is, therefore, given by
the
following functional determinant on the space of functions of three
spatial
coordinates
	\begin{eqnarray}
	\mbox{\boldmath $\Gamma$}_\beta
        &=& - \ln \left[  \int{D \varphi_{-}}< \varphi _{-}\mid
	\exp{(- \beta  \hat{H})} \mid \varphi _{-} >  \right]  \\
	&=& -{1 \over 2}
	\ln \det \left[ \ {1 \over {2(\cosh{\beta\hat{\omega}} -1)}}
	\delta ({\mbox{\boldmath $x$}}-{\mbox{\boldmath $y$}}) \
\right]  .
        \label{G}
	\end{eqnarray}

It is worth emphasizing that all the quantities and operators
entering the WKB
approximation of the wave function and density matrix depend  on  a
3-geometry
of space and values of fields on it. The whole information about
4-dimensional
manifold is contained in  the interval $\beta$ of Euclidean time
between the
points with the same spatial coordinates ${\mbox{\boldmath $x$}}$of
spacelike
slices
${\mbox{\boldmath $\Sigma_+$}}$ and ${\mbox{\boldmath $\Sigma_-$}}$.
In the
case of the Schwarzschild black hole $\beta=8\pi M$.

The density matrix $\hat{\rho_\beta}$ satisfies the equation
	\begin{eqnarray}
	{\partial {\hat{\rho}_{\beta}}\over \partial \beta} =
	({\partial \mbox{\boldmath $\Gamma$}_\beta \over \partial
\beta } - \hat{H} )
        \hat{\rho}_{\beta}  .
	\end{eqnarray}
Using this relation one can easily show that the entropy of the
system in
question can be obtained from the effective action $\mbox{\boldmath
$\Gamma$}_\beta$
	\begin{eqnarray}
	&&\mbox{\boldmath $S$}=\mbox{\boldmath $S$}_\beta
\Big|_{\,\beta=8\pi M}   \ \
 , \\
	&&\mbox{\boldmath $S$}_\beta \equiv
-\mbox{Tr}[\,\hat\rho_\beta \ln{\hat
\rho_\beta}\, ]  =
	-\mbox{Tr}[ \,(\mbox{\boldmath $\Gamma$}_\beta - \beta\hat{H}
)
        \hat{\rho_\beta}\,]
	=  \beta  {\partial \mbox{\boldmath $\Gamma$}_\beta\over
\partial
        \beta}-\mbox{\boldmath $\Gamma$}_\beta.  \label{5.18}
	\end{eqnarray}
Note that it would be incorrect to differentiate directly
$\mbox{\boldmath
$\Gamma$}$ over $M$ in order to obtain the entropy $\mbox{\boldmath
$S$}$,
since the total effective action is an integral over the whole space
and
depends also on its geometry. The Hawking temperature
$T_{BH}=1/8\pi M$ depends both on the space-geometry and on
$g_{\tau\tau}$ of
the four-dimensional metric and hence operations of differentiation
over $M$
and integration over volume do not commute in general case. In order
to avoid
this difficulty we introduced the generalized density matrix
$\hat{\rho_\beta}$.

      In order to calculate $\mbox{Tr}\ln$ entering  the expression
for the
effective action, it is convenient to expand all the functions
$\varphi
({\mbox{\boldmath $x$}})$ in terms of eigenfunctions $R_\lambda
({\mbox{\boldmath $x$}})$ of the operator $\hat{\omega}$
	\begin{eqnarray}
	\varphi ({\mbox{\boldmath $x$}})=
        \sum_{\lambda} {\varphi _\lambda R_\lambda ({\mbox{\boldmath
$x$}})},
	\hskip 2 cm
	\hat{\omega}^2 R_\lambda ({\mbox{\boldmath $x$}}) =
        \omega_\lambda ^2 R_\lambda ({\mbox{\boldmath $x$}}) ,
	\end{eqnarray}
	\begin{eqnarray}
	\delta ({\mbox{\boldmath $x$}}-{\mbox{\boldmath $y$}})=
        \sum_{\lambda} g^{\tau\tau} g^{1 \over 2}
	R_\lambda ({\mbox{\boldmath $x$}}) R_\lambda
({\mbox{\boldmath $y$}}),
	\end{eqnarray}
Here $  \sum_{\lambda} $ denotes the sum over all quantum numbers
$\lambda$.
Substitution of the expansion of $\delta$-function in terms of
eigenfunctions of the operator $\hat{\omega}$ gives
	\begin{eqnarray*}
	\mbox{\boldmath $S$}_\beta&=&\int {d {\mbox{\boldmath $x$}}}
        \left( \beta{\partial \over \partial \beta}
	- 1 \right)\left[ \ln{(2 \sinh{{\beta \over 2}
        \hat{\omega_x}})}
	\delta ({\mbox{\boldmath $x$}} - {\mbox{\boldmath $y$}})
\right]
        _{{\mbox{\boldmath $y$}}= {\mbox{\boldmath $x$}}} \\
	&=&\int {d {\mbox{\boldmath $x$}}} \left[
	{\beta \over 2}\hat{\omega}_y
	\coth{{\beta \over 2}\hat{\omega}_y}  -
	\ln{(2 \sinh{{\beta \over 2}\hat{\omega}_y})}
	 \right]     \times \\
	&&\hskip 4cm  \sum_{\lambda} \left(
 	 g^{\tau\tau}({\mbox{\boldmath $x$}}) g ^{1 \over 2}
({\mbox{\boldmath $x$}})
	 R_\lambda ({\mbox{\boldmath $x$}})
	 R_\lambda ({\mbox{\boldmath $y$}})
	 \right) _{{\mbox{\boldmath $y$}}= {\mbox{\boldmath $x$}}}
\\
	&=& \int {d {\mbox{\boldmath $x$}}} g^{\tau\tau} g ^{1 \over
2} \sum_{\lambda}
	R_\lambda ({\mbox{\boldmath $x$}}) ^2 \left[
	{\beta \over 2} \omega_\lambda  \coth{{\beta \over 2}
\omega_\lambda } -
	\ln{(2 \sinh{{\beta \over 2} \omega_\lambda })}
	\right]  .
	\end{eqnarray*}
Thus we have
	\begin{eqnarray}
	\mbox{\boldmath $S$}_\beta= \int {d {\mbox{\boldmath $x$}}}
        \sum_{\lambda}  \mu _\lambda  ({\mbox{\boldmath $x$}})
	s (\beta\omega_\lambda ) ,
	\end{eqnarray}
Here
	\begin{eqnarray}
	s (\beta\omega ) = {{\beta\omega } \over {e^{\beta\omega
}-1}} -
	\ln (1- e^{-\beta\omega })
	\end{eqnarray}
is a well known expression for the entropy of single oscillator with
the
frequency $\omega$ at temperature $T=1/\beta$ and
	\begin{eqnarray}
	\mu _\lambda  ({\mbox{\boldmath $x$}}) = g^{\tau\tau} g ^{1
\over 2}
	R_\lambda ({\mbox{\boldmath $x$}}) ^2
	\end{eqnarray}
is a  phase space density of quantum modes. In order to estimate the
contribution of regions of space in the vicinity of
the horizon into the entropy of black hole we should  find an
asymptotic
solution for the mode functions $R_\lambda ({\mbox{\boldmath $x$}})$
near the
horizon. Eigen functions $R_\lambda ({\mbox{\boldmath $x$}})$ for a
massless
scalar field in the Schwarzschild spacetimeare of the
form
	\begin{eqnarray}
	R_\lambda (r,\Omega) =  R_{\omega l}(r)Y^l _{m}(\Omega)  .
	\end{eqnarray}
Here $Y^l _{m}(\Omega) = {1 \over \sqrt{2\pi}} e^{i m \varphi} P^m_l
(\cos
\theta) $ are spherical functions and  radial functions $R_{\omega
l}(r)$ are
real and obey the equation
	\begin{eqnarray}
	\left[ {d \over {d r}}(r^2 - 2 M r) {d \over {d r}} - l ( l+1
) + \omega^2
	 {{r^3} \over {r -2 M}}  \right]  R_{\omega l}( r ) = 0.
\label{5.29}
	\end{eqnarray}
The expression for an entropy of such a system takes the form
	\begin{eqnarray}
	\mbox{\boldmath $S$} = \int_{2 M}^{\infty} d r r^2 (1-{{2 M}
\over r})^{-1}
	 \int_{0}^{\infty} d\omega \sum_{l=0}^{\infty} (2 l+1)
R_{\omega l}(r) ^2
	s (8 \pi M \omega) .
	\label{entropy}
	\end{eqnarray}
Regular near the horizon solutions of this equation  are normalized
by the
condition
	\begin{eqnarray}
	 \int_{2 M}^{\infty} d r r^2 (1-{{2 M} \over r})^{-1}
         R_{\omega l}(r) R_{\omega ' l}(r) = \delta
(\omega-\omega ').
	\end{eqnarray}
The entropy of an oscillator  with frequency $\omega$  exponentially
decreases
for frequencies much larger than the  black hole temperature $ T= 1/
8\pi M $.
 In the vicinity of the horizon $\mid \xi-1 \mid \ll 1$ a regular
solution for
radial modes takes a simple form
	\begin{eqnarray}
        R_{\omega  l}(r) \simeq A(M, \omega, l)
        K_{i 4 M \omega} \left( \sqrt{ 2 l (l+1)( {r \over 	M} -
2) }  \right).
	\end{eqnarray}
The normalization factor $A(M, \omega, l)$ depends on the coefficient
of
penetration of modes through the potential barrier. All the modes  in
the range
of frequencies in question and with  angular quantum numbers $l  \geq
3 $ are
trapped. For such modes  the penetration coefficient is exponentially
small and
normalization factor  does not depend on $ l $ . The larger $l$ the
closer to
the horizon a return point lies  and the better approximation
becomes.
The evaluation of normalization factor gives
	\begin{eqnarray}
	A(M, \omega, l)  \simeq \left[ {{2 \omega  \sinh{4 \pi M
\omega}} \over {M
\pi^2}} \right] ^{1 \over 2},
	\end{eqnarray}
where the relation
	\begin{eqnarray}
	\int _0 ^\infty dy \ {1 \over y} \  K _{ix}(y) K _{ix'}(y) =
	{{\pi ^2} \over 2} {1 \over {x \sinh (\pi x) }} \delta (x-x')
	\end{eqnarray}
was used. We also use the fact that the modified Bessel  functions
decrease
very fast with increasing of their argument and, hence, with a good
accuracy
the integration along radius can be extended to infinity.
One can see that the main contribution  to the integral of entropy
near the
horizon comes from large  $l$. Replacement in Eq.(\ref{entropy}) of
summation
over $l$ by integration leads to the expression
	\begin{eqnarray}
	\sum_{l=0}^{\infty} (2 l+1) R_{\omega l}(r) ^2
         \simeq 2 \int_{0}^{\infty} d l  l  R_{\omega l}(r) 	^2 =
{{4 \omega^2}
\over \pi}{M \over {r-2M}}.
	\end{eqnarray}
This asymptotic formula  reproduces the result by Candelas and Howard
\cite{Candelas} for a mode summation in Schwarzschild geometry.
For evaluation of the integral (\ref{entropy}) it is convenient to
split the
region of integration into two points. In first part $2 M \leq r \leq
r_0 = 3
M$ we can use the above described approximation. Namely this
contribution is
related to the entropy of a black hole $\mbox{\boldmath $S$}_H$. The
integration over another region $(r \leq 3 M)$ formally diverges at
$r
\rightarrow \infty$. This divergence is simply connected with the
fact that we
consider black hole in equilibrium with infinite reservoir of thermal
radiation. But this equilibrium is unstable. In order to get stable
equilibrium
one needs to insert a black hole into a cavity of the size comparable
with
$r\approx3M$. For such physical problem the second contribution
(related to the
entropy of a thermal gas  far from the black hole) becomes negligibly
small and
we can simply omit this term \cite{FrNo:93a,FrNo:93b}. That is why we
have
	\begin{eqnarray}
	\mbox{\boldmath $S$}_H &=& {{4 M} \over \pi}
	\int_{2 M} ^{r_0} {d r {{r^3} \over (r-2M)^2}}
 	\int_{0}^{\infty}{d \omega  \omega^2 s(8 \pi M \omega)}
\nonumber \\
	&\simeq& {{512 M^5} \over \pi} \int_{0}
        ^{z_0} d z{1 \over z^3}
	 \int_{0}^{\infty}{d \omega  \omega^2 s(8 \pi M \omega)}
\label{S} \\
	&=& {{ 4 M^2} \over 45} \int_{0}
        ^{z_0}d z{1 \over z^3} \ \ ,
	 \nonumber
	\end{eqnarray}
where $z_0= r_0 \sqrt{1-{2M \over r_0}} + M \ln [ \ {r_0 \over M}  -1
+  {r_0
\over M}
 \sqrt{1-{2M \over r_0}} \ ] $ is a proper distance from the horizon
to the
point $r_0$. This result shows that one-loop  contribution to the
entropy of
black hole $\mbox{\boldmath $S$}_H$ diverges near the horizon.The
expression
(\ref{S}) gives the leading divergent term  and reproduces the result
by Frolov
and Novikov \cite{FrNo:93a,FrNo:93b}. This divergence is physical and
its
origin does not depend on particular properties of quantum fields
surrounding a
black hole. The analogous divergence evidently occurs for higher
spins. Hence
quantum corrections can never be neglected in description of
thermodynamical
properties of black holes. It is worth to emphasize that  shifting of
a
position of the horizon as a whole due to the back reaction effect of
quantum
fields on the geometry of black hole does  not remove the divergence.
Fluctuations of the horizon are to be taken into account to provide
the
necessary cutoff.

\section{Entropy and Effective Action}
\hspace{\parindent}
In previous section we used the proposal for a wave function of black
hole in
the calculation of an entropy of a scalar field in the vicinity of
Schwarzschild black hole. Only the properties of 3-dimensional space
and
fields on it were used  in the consideration. It would be interesting
to
compare  this result with that of  the 4-dimensional Euclidean action
approach. This also allows one to generalize the result of section 5
to
arbitrary static black holes.

Consider  4-dimensional Euclidean effective action $\mbox{\boldmath
$\Gamma$}_\beta$ for a conformal scalar field
$\phi(\tau,{\mbox{\boldmath
$x$}})$ with Hamiltonian Eq.(\ref{H}) on a manifold periodic in
Euclidean time
with period $\beta$. Up to a contribution of a local functional
measure  it can
be represented  in the form
      \begin{eqnarray}
      \mbox{\boldmath $\Gamma$}_\beta
      &=& + {1 \over 2} \mbox{Tr}\ln  F + \delta^4 (0)  (\dots) \ \ ,
      \label{G4} \\
      F&=& - \Box + {1 \over 6} R \ \ .
      \end{eqnarray}
     This effective action and the corresponding free energy  $
\mbox{\boldmath
$F$}_\beta=\mbox{\boldmath $\Gamma$}_\beta/ \beta $ have ultraviolet
divergences. Note that, though the last term in Eq.(\ref{G4})
diverges,  it is
 proportional  to $\beta$ and hence the  free energy does not depend
on $\beta$
and its contribution into entropy vanishes.
The same argument remains valid for all ultraviolet divergences in
the
effective action. Thus we have
     \begin{eqnarray}
     \mbox{\boldmath $S$}=\beta^2 {\partial \over \partial\beta} \
\mbox{\boldmath $F$}_\beta =\beta^2
     {\partial \over  \partial\beta} \  \mbox{\boldmath $F$}_\beta
^{\mbox{\scriptsize Ren}}.
     \end{eqnarray}
The effective action and thermodynamic potential for scalar fields at
finite
temperature in static curved spacetime were calculated by Dowker and
Schofield
\cite{Dowker}. It was proved  that in the case of the conformal
scalar field
the effective actions in two conformally related spaces
$\bar{g}_{\mu\nu}=e^{-2
\omega} g_{\mu\nu}$ are related to each other by the equation
          \begin{eqnarray}
         \Delta {\mbox{\boldmath $\Gamma$}} [g,\omega] &=&
         {\mbox{\boldmath $\Gamma$}}_\beta ^{\mbox{\scriptsize Ren}}
          [e^{\bar{g}}] -
         \mbox{\boldmath $\Gamma$}_\beta^{\mbox{\scriptsize Ren}} [g]
\ ,
         \nonumber \\
         \Delta {\mbox{\boldmath $\Gamma$}} [g,\omega] &=&
         - {1 \over 2880 \pi^2} \int_{0}^{\beta}d \tau \int{g^{1
\over 2}}
d^{3} x
         \left[
        + 3 (\Box \omega)^2
        - 4 \omega^\sigma \omega_\sigma \Box \omega  \right.
\nonumber \\
        &+& 2 (\omega^\sigma \omega_\sigma)^2
        - 2 R_{\mu\nu} \omega^\mu \omega^\nu   \label{Ups}
        + \left. \omega \left\{  R_{\alpha\beta\gamma\delta}
        R^{\alpha\beta\gamma\delta} - R_{\alpha\beta}
R^{\alpha\beta}+
        \Box R   \right\}
        \right] \  ,  \\
        \mbox{\boldmath $\Gamma$}[\bar{g}]&=&{1 \over 2} \mbox{Tr}\ln
\bar{F}
\ ,
        \hskip 1.5cm
        \bar{F} = - \stackrel{-}{\Box} + {1 \over 6}\bar{R} \ ,
        \hskip 1.5cm
        \bar{g}_{\mu\nu}=e^{-2\omega}g_{\mu\nu}  \ .  \nonumber
        \end{eqnarray}
The difference $\Delta {\mbox{\boldmath $\Gamma$}} [g,\omega]$ for
two
conformally related theories is proportional to $\beta$ and hence
does not
contribute to the entropy. We apply these relations to the particular
case of
an ultrastatic metric $\bar{g}$ , i.e. when  $\omega\equiv{1 \over
2}\ln
g_{\tau\tau}$ and $\omega_\mu = \nabla_\mu \omega$.

For the calculation of  an  effective action in the ultrastatic space
$\bar{g}$
it is convenient to apply  the heat-kernel technique.
In a proper time representation, the effective action of a scalar
field on  an
ultrastatic 4-dimensional Euclidean manifold
      \begin{eqnarray}
      d\bar{s}^2 &=& d \tau^2 + d\bar{l}^2 \label{bars}  \ , \\
      d\bar{l}^2 &=& {1 \over (1- 2M/r )^2} dr^2 +
      {r^2 \over 1- 2M/r} (d\theta^2 + \sin ^2 \theta  d\varphi^2) \
,
\nonumber
      \end{eqnarray}
 which is periodic in $\tau$ with constant period $\beta$ takes the
form
     \begin{eqnarray}
     \mbox{\boldmath $\Gamma$} [\bar{g}]
     \equiv \bar{\mbox{\boldmath $\Gamma$}}_\beta  &=&
     - {1 \over 2}\int_{0}^{\infty}{{d s} \over s}
     \mbox{Tr}\bar{K}_\beta (s) \ .
     \end{eqnarray}
The heat kernel  $ \bar{K}_\beta $  periodic in $\tau$ with a period
$\beta$
is a solution of the problem
     \begin{eqnarray}
     {\partial \over \partial s}{\bar{K} _\beta  (s \vert
\tau,{\mbox{\boldmath
$x$}}
     ; \tau ' ,{\mbox{\boldmath $x$}}')} &=&
     \bar{F}
     \bar{K} _\beta (s \vert \tau,{\mbox{\boldmath $x$}} ;
      \tau ' ,{\mbox{\boldmath $x$}}')  \ , \nonumber \\
      \bar{K} _{\beta} (s \vert \tau,{\mbox{\boldmath $x$}} ;
     \tau ' ,{\mbox{\boldmath $x$}}')
     &=& \bar{K}_\beta (s \vert \tau + \beta,{\mbox{\boldmath $x$}} ;
      \tau '  ,{\mbox{\boldmath $x$}}') \ , \\
      \bar{K} _{\beta} (0 \vert \tau,{\mbox{\boldmath $x$}} ;
     \tau ' ,{\mbox{\boldmath $x$}}')
     &=& \delta (\tau-\tau ')  \delta
     ({\mbox{\boldmath $x$}}-{\mbox{\boldmath $x$}}') \ .  \nonumber
     \end{eqnarray}
It can be obtained by the method of images
      \begin{eqnarray}
      \bar{K} _{\beta} (s \vert \tau,{\mbox{\boldmath $x$}} ;
      \tau ' ,{\mbox{\boldmath $x$}}')&=& \sum _{n=-\infty} ^\infty
      (s \vert \tau,{\mbox{\boldmath $x$}} ;
      \tau '+ \beta n ,{\mbox{\boldmath $x$}}')
      \label{KK}
      \end{eqnarray}
from the nonperiodic heat kernel $ \bar{K}=\bar{K}_\infty $ defined
on  a
complete interval $- \infty < \tau, \tau' <\infty $.
Due to the separation of variables in the operator
$\bar{F}=-\partial^2 /
\partial \tau^2 -\bar{\Delta} + 1 /6 \bar{R}$, the heat kernel $
\bar{K} $
takes the form
      \begin{eqnarray}
      \bar{K} (s \vert \tau,{\mbox{\boldmath $x$}} ;
      \tau ' ,{\mbox{\boldmath $x$}}')=(4 \pi s)^{-{1 \over 2}}
      \exp \left[ -{(\tau-\tau ')^2 \over 4 s} \right] {}^3\!\bar{K}
     (s \vert {\mbox{\boldmath $x$}} ;{\mbox{\boldmath $x$}}') \ \ ,
      \label{K}
      \end{eqnarray}
where ${}^3\!\bar{K}(s \vert {\mbox{\boldmath $x$}} ;{\mbox{\boldmath
$x$}}')$
is 3-dimensional analogue of the heat kernel corresponding to the
operator $
- \bar{\Delta} + 1 /6 \bar{R} $ . From Eq.(\ref{KK}) and Eq.(\ref{K})
we have
      \begin{eqnarray}
      \bar{K}_\beta (s \vert \tau,{\mbox{\boldmath $x$}} ;
      \tau ' ,{\mbox{\boldmath $x$}}') =
      \theta_3 \left(-i {(\tau-\tau ')\beta \over 4\pi s},
      \exp \left[ -{\beta^2 \over 4 s} \right]  \right)
      \bar{K} (s \vert \tau,{\mbox{\boldmath $x$}} ;
      \tau ' ,{\mbox{\boldmath $x$}}')\ ,
      \end{eqnarray}
where  $\theta_3$ is a Riemann theta function
       \begin{eqnarray}
       \theta _3 (0, \exp [-b]) = \sum _{n=-\infty} ^\infty \exp [ -b
n^2 ]  \
{}.
       \end{eqnarray}

The "zero-temperature" heat kernel $\bar{K} (s \vert
\tau,{\mbox{\boldmath
$x$}} ; \tau ' ,{\mbox{\boldmath $x$}}')$ can be expanded in nonlocal
series in
powers of curvatures of a spacetime \cite{Zelnikov,CPT}. To calculate
the
effective action we need to know the trace of the heat kernel with
coincident
points $ (\tau = \tau', \ {\mbox{\boldmath $x$}} = {\mbox{\boldmath
$x$}}' )$.
In the notations of  \cite{CPT}   it reads
      \begin{eqnarray}
      \mbox{Tr}\bar{K}_\beta (s \vert \tau,{\mbox{\boldmath $x$}} ;
      \tau  ,{\mbox{\boldmath $x$}}) =
      \theta_3 \left(0,\exp \left[ -{\beta^2 \over 4 s} \right]
\right)
      \mbox{Tr}\bar{K} (s \vert \tau,{\mbox{\boldmath $x$}} ;
      \tau ,{\mbox{\boldmath $x$}}) \ \ ,
      \label{K1}
        \end{eqnarray}
        \begin{eqnarray}
      \mbox{Tr}\bar{K} (s \vert \tau,{\mbox{\boldmath $x$}} ;
      \tau ,{\mbox{\boldmath $x$}})&=&{1 \over (4 \pi s)^2}
      \int_{0}^{\beta}{d\tau}\int{d^3 x \sqrt{\bar{g}}}
      \left\{
      1 + s \bar{P} \right. \nonumber \\
       &+&\left.  s^2 \left[
       \bar{R}_{\mu\nu} f_1(-s\stackrel{-}{\Box}) \bar{R}^{\mu\nu}
       +  \bar{R} f_2(-s\stackrel{-}{\Box}) \bar{R}   \right.\right.
\\
       &+& \left.\left. \bar{P} f_3(-s\stackrel{-}{\Box}) \bar{R}
       +  \bar{P} f_4(-s\stackrel{-}{\Box}) \bar{P}
       \right]
       \right\}  \nonumber  \\
       &+& O(\mbox{Curvatures} { }^3) \ \ ,  \nonumber
       \label{UltraW}
         \end{eqnarray}
where $f_i (-s\stackrel{-}{\Box}) $  are  nonlocal form factors
\cite{CPT} and
$\bar{P}=0$ for conformal scalar field.
The first two terms in this expression are local. Nonlocalities
appear only in
quadratic and in higher orders in curvature terms.

We apply this formula to the Schwarzschild geometry.
Near the horizon the corresponding ultrastatic metric Eq.(\ref{bars})
with a
good accuracy describes the geometry of $R^1 \times H^3$ space. Where
$H^3$ is
a space of constant negative curvature
      \begin{eqnarray}
      ds_H^2 = d\tau^2 + d l^2 + (4M)^2 \sinh^2 \left[{ l\over 4M}
\right]
     (d \theta^2 + \sin^2 \theta d \varphi^2)\ \ .
       \end{eqnarray}
The difference between this metric and the metric (\ref{bars})
     \begin{eqnarray}
     h^\mu_\nu = \bar{g}^{\nu\sigma}[\bar{g}_{\nu\sigma} - g_{H
\mu\sigma}] =
     \Delta(r)\mbox{diag} (0,0,1,1)
     \end{eqnarray}
     \begin{eqnarray}
     \Delta(r) = 6 ({r / 2M}-1) + O\left( ({r / 2M}-1)^2 \right)
     \end{eqnarray}
vanishes at the horizon as $ \sim r - 2M $. One can show that local
invariants
constructed from the tensors of the form $\nabla_{\alpha
\dots}\nabla_\beta
\nabla ^{\gamma \dots}\nabla^\delta  h^\mu_\nu = O(r-2M)$ vanish at
the horizon
too. That is why all the invariants $\bar{I}$ constructed from the
metric
$\bar{g}$ and curvature $\bar{R}$ differ from the corresponding
invariants
$I_H$
for the metric $g_H$ by terms vanishing at the horizon $I=I_H
+O(r-2M)$.

The heat kernel on spaces of constant negative curvature is known
explicitly
and for a conformally invariant scalar field it reads
\cite{Camporesi}
      \begin{eqnarray}
       \bar{K}_H (s \vert \tau,{\mbox{\boldmath $x$}} ;
       \tau' ,{\mbox{\boldmath $y$}}) = {1 \over (4 \pi s)^2}
      {\sigma({\mbox{\boldmath $x,y$}})
	\over 4M \sinh\,[\,\sigma({\mbox{\boldmath $x,y$}}) / 4M\,]}
       \exp\left[{\tau^2+\sigma ^2 ({\mbox{\boldmath $x,y$}}) \over
4s}\right]
      \sqrt{g_H}\ \ .
      \end{eqnarray}
Here $ \sigma({\mbox{\boldmath $x,y$}})$ is the geodesic distance on
the $H^3$
space section.

The relation between $K$ and $K_H$  for coincident points implies $
\bar{K} (s
\vert \tau,{\mbox{\boldmath $x$}} ; \tau ,{\mbox{\boldmath $x$}})={1
/ (4 \pi
s)^2} + O(r-2M)$ and we have
       \begin{eqnarray}
       \mbox{Tr}\bar{K}_\beta (s \vert \tau,{\mbox{\boldmath $x$}} ;
        \tau ,{\mbox{\boldmath $x$}})
       ={1 \over (4 \pi s)^2}
       \theta_3 \left(0,\exp \left[ -{\beta^2 \over 4 s} \right]
\right)
       \int_{0}^{\beta}{d\tau}\int{d^3 x \sqrt{\bar{g}}} [1+O(r-2M)]
\ \ .
       \end{eqnarray}
Composition of this expression, Eq.(\ref{Ups}) and Eq.(\ref{K1})
gives for the
free energy of a conformal scalar field the expression
        \begin{eqnarray}
        { \mbox{\boldmath $F$}_\beta^{\mbox{\scriptsize Ren}}}  -
	\mbox{\boldmath $F$}_\infty^{\mbox{\scriptsize Ren}} =
        -  {\pi^2 \over 90}  \int d {\mbox{\boldmath $x$}}
	g ^{1 \over 2}
        \left[  (g_{\tau\tau})^{-2}  {1 \over \beta^4}  -
	\left( {1 \over 2\pi} \right)^4
        \left( \omega^\sigma  \omega_\sigma \right)^2
        \right] + \dots    \ \ ,
 \label{6.19}
        \end{eqnarray}
where the integral relation
        \begin{eqnarray}
        \int_{0}^{\infty}dx x^{a-1} \left[ \theta_3(0,e^{-x}) -1
\right] =
        2\Gamma(a) \zeta(2 a) \ \
        \stackrel{\mbox{\scriptsize for} \  a=2}{=} \ \  {\pi^4 \over
45}
        \end{eqnarray}
has been used and the physical metric $g_{\mu\nu}$ has been restored.
Dots
designate terms which are less divergent or finite at the horizon.
The
corresponding entropy
     \begin{eqnarray}
     \mbox{\boldmath $S$} &=& \beta^2{\partial \over \partial \beta}
     { \mbox{\boldmath $F$}_\beta^{\mbox{\scriptsize Ren}}} =
     \beta^2{\partial \over \partial \beta} \left[ { \mbox{\boldmath
$F$}_\beta^{\mbox{\scriptsize Ren}}}
      -  \mbox{\boldmath $F$}_\infty^{\mbox{\scriptsize Ren}}
\right]
\nonumber \\
     &=&  {2 \pi^2 \over 45 }{1 \over \beta^3}
      \int d {\mbox{\boldmath $x$}} (g_{\tau\tau})^{-2}
      g ^{1 \over 2} +  \dots    \ \ ,
            \label{6.21}
      \end{eqnarray}
calculated  \ for a particular case of \ Schwarzschild black hole \
reproduces
the result Eq.(\ref{S}).

It is worth emphasizing that as a result of the restoration of the
physical
metric
       \begin{eqnarray}
       \bar{g}_{\mu\nu}&=&
       e^{-2\omega}g_{\mu\nu} = {1 \over g_{\tau\tau}}g_{\mu\nu} \ ,
       \hskip 2cm
       \sqrt{\bar{g}} = e^{-4\omega} \sqrt{g} = {1 \over
(g_{\tau\tau})^2}
\sqrt{g} \ , \\
       \bar{R}_{\mu}^{\nu} &=& e^{2\omega} \left[
       R_{\mu}^{\nu} + 2 \omega_\mu^{;\nu} + \Box \omega \delta
_{\mu}^{\nu}
       + 2 \omega_\mu \omega^\nu
       - 2 \omega^\sigma  \omega_\sigma \delta _{\mu}^{\nu}
       \right] \nonumber
       \end{eqnarray}
in the general expansion  Eq.(\ref{K1}) we get a nonlocal expansion
of the
effective action in terms of curvature,  "acceleration" $\omega_\mu$
and their
derivatives. One can use this effective action in order to get
$<T_{\mu\nu}>^{\mbox{\scriptsize Ren}}$. The action can be written in
a
completely invariant form if we substitute $g_{\tau\tau} = g_{\mu\nu}
\xi^\mu
\xi^\nu$ and consider $\xi^\mu$ as external field, which is fixed
during the
variations over $g_{\mu\nu}$ and is taken to coinside with the
Killing vector
field after the variations were performed. An additional (external)
vector
field $\xi$ in the effective action for thermal state is required
because such
a state is possible only in a stationary spacetime, i.e. the
spacetime with
additional geometric structure.

\section{Conclusions}
\hspace{\parindent}
In conclusion we make some  remarks concerning the obtained result.
The
proposed no-boundary anzatz for a wave function of a black hole
appears to be a
natural approach, which in particular allows one to give a covariant
description of its degrees of freedom. The quantum state of a black
hole is
characterized by an amplitude of different realizations of dynamical
variables
on the Einstein-Rosen bridge. For  any particular realization one can
define a
Euclidean horizon by means of the following procedure. Take a
two-sphere $S$
from the first non-trivial homotopic class $\pi_2$ and define $A[S]$
as the
surface area of $S$. Because $S$ is non-contractable the functional
$A[S]$ has
a non-vanishing minimum. The corresponding sphere $S_0$ defines a
position of
the Euclidean horizon for a chosen realization. A surface $S_0$
changes from
one realization to another. This dependence of $S_0$ on the
realization can be
interpreted as quantum fluctuations of the horizon. The effect of
quantum
fluctuation ({\em zitterbewegung}) of the horizon is important for
the problem
of entropy discussed in the paper.

The one-loop calculations of entropy of internal degrees of freedom
of a black
hole was shown to be divergent at the horizon. The divergency has a
universal
law near the horizon of a black hole for all fields (massless and
massive, with
and without spin).  The divergences  arise because in our one-loop
approximation the background geometry (and hence the position of the
Euclidean
horizon) is fixed. Quantum fluctuations of the horizon result in its
spreading.
Due to spreading we cannot any more to split the states of quantum
fields
located inside the region of a fluctuating horizon into the 'visible'
and
'invisible' ones. In other words for any chosen realization of a
quantum field
the splitting of states into internal and external states of a black
hole
depends on the realization. Averaging over different realization
(which
effectively takes into account the zitterbewegung of the horizon) may
produce
the required cut-off for the entropy.

It is expected that the quantum effects with the properly described
fluctuations of the horizon must give the standard expression $A/4$
for the
entropy of a black hole. It is important that this result must not
depend on
the number of fields and their properties. We should emphasize that
in the
framework of our approach the dynamical degrees of freedom of a black
hole
contribute to the entropy only on the one-loop level, while there is
no
tree-level contributions. The remarkable fact is that in the standart
Euclidean
action approach the "correct" answer for the entropy $(A/ 4 l^2_P)$
is obtained
by calculating the topological tree-level contribution into the
Euclidean
gravitational action. The relation between "dynamical" and
"topological"
contributions to the entropy as well as the origin of the
universality of the
expression for the entropy of a black hole is a real puzzle.

       Recently, an elegant proposal \cite{SuUg:94} has been given
for a
mechanism maintaining the exact relation between the black hole
entropy and its
horizon area on the nonperturbative level of quantum gravitational
thermodynamics in the limit of very heavy black holes.  Briefly it
looks as
follows. Suppose, we have the gravitational effective action of the
theory
$\mbox{\boldmath $\Gamma$}[\,g\,]$, possibly generated by the
fundamental
theory of (super)strings and , therefore, finite. It may have a very
general
structure about which only one assumption is made: it is supposed to
be
analytic in the curvature and free from the effective cosmological
term (thus
admitting the existence of the asymptotically flat solutions of
effective
Einstein equations)
	\begin{eqnarray}
	\mbox{\boldmath $\Gamma$}[\,g\,]=\sum_{n=1}^{\infty}\,\int
dx_1...dx_n\,\mbox{\boldmath $\Gamma$}_{n}
	(x_1,...x_n)\,R(x_1)...R(x_n).
\label{7.1}
	\end{eqnarray}
Here $R(x)$ is a collective notation for the curvature and Ricci
tensors and
 $\mbox{\boldmath $\Gamma$}_{n}(x_1,...x_n)$ is a set of (generally
nonlocal)
form factors
accumulating all the information about the quantum and statistical
effects in
the theory. Since these form factors represent the coordinate kernels
of some
nonlocal operators constructed of derivatives, the only covariant
expression
available for $\mbox{\boldmath $\Gamma$}_1(x)$ is just the local
density
	\begin{eqnarray}
	\mbox{\boldmath $\Gamma$}_1(x)=-\frac1{16\pi^2 l_{\rm
eff}^2}\,g^{1/2}(x)
    \label{7.2}
	\end{eqnarray}
with a purely numerical coefficient which can be identified with the
effective
(renormalized) gravitational constant or Planck length $l_{\rm eff}$
(all the
covariant derivatives in $\mbox{\boldmath $\Gamma$}_1$ contract to
form a total
derivative which disappears when integrated over asymptotically flat
spacetime).

According to eq. (\ref{5.18}) the calculation of entropy involves the
effective
action $\mbox{\boldmath $\Gamma$}_{\beta}=\mbox{\boldmath
$\Gamma$}[\,g^{\beta}\,]$ calculated on the conical spacetime with
metric
$g^{\beta}$ having a conical singularity with $\beta\not=8\pi M$.  On
such a
manifold the curvature has a form
	\begin{eqnarray}
	R_{\beta}(x)=(\beta-8\pi M)\,f(x)+R_{\rm reg}(x),
\label{7.3}
	\end{eqnarray}
where $R_{\rm reg}(x)$ is a regular part of the curvature bounded by
$1/M^2$
and, therefore, negligible for heavy black holes
$M\rightarrow\infty$. The
singular part caused by conical structure for $\beta\not=8\pi M$
involves the
generalized function $f(x)$ which, when regulated, can be even
nonsingular one,
but having the compact support in the vicinity of the tip of the cone
(black
hole horizon) and satisfying the relation
	\begin{eqnarray}
	\int dx\,g^{1/2}(x)\,f(x)=-8\pi M \ .
\label{7.4}
	\end{eqnarray}
Substituting the structure (\ref{7.3}) into (\ref{7.1}) and using
(\ref{5.18})
we immediately find that the entropy is entirely generated by the
effective
Einstein term of the action, because the expansion in powers of the
curvature
becomes the expansion in powers of the angle deficit $(\beta-8\pi M)$
of the
conical manifold:
	\begin{eqnarray}
	\mbox{\boldmath $S$}=\left(
\beta\frac{\partial}{\partial\beta}
      -1\right)\,\mbox{\boldmath $\Gamma$}_{\beta}=
	\beta\int dx \,\mbox{\boldmath $\Gamma$}_1(x)\,R(x)
      =\frac{A}{4\,l_{\rm eff}^2} \ .     \label{7.5}
	\end{eqnarray}

The above arguments could have been even generalized to the case of
the
finite-mass black hole by noting that in asymptotically flat
spacetime the
actual expansion of the effective action can be performed in powers
of the
Ricci curvature $R_{\mu\nu}$ only \cite{CPT,nonloc}, for which
$R_{\mu\nu\,\rm
reg}(x)\equiv 0$ in eq.(\ref{7.3}). However there is a serious
objection to
this mechanism which apparently invalidates this proposal.  If it
were correct
then the perturbative calculations of entropy would maintain the
universal
relation between the entropy and one quarter of the horizon area in
units of
the effective Planck length, the quantum corrections to the classical
entropy
being compensated by the simultaneous renormalization of this length.
But this
is definitely not the case for the dominant divergent contribution
(\ref{6.21})
obtained in the one-loop approximation. Indeed, as it follows from
eq.(\ref{6.19}),  this contribution involves the invariant of the
Killing
vector field $xi^\mu \xi_\mu=g_{\tau\tau}$. This invariant can be
regarded as a
restriction of some nonlocal functional of metric to the manifold
with Killing
symmetries. Killing field $\xi^{\mu}$ as a functional of the metric
does not
have a unique continuation off the symmetric (Killing) points in the
configuration space of metric, but it is undoubtedly nonlocal and
most likely
has a structure of the solution of the Killing equation
	\begin{eqnarray}
	\Box \xi^{\mu}+R^\mu_\nu\,\xi^{\nu}=0        \label{7.6}
	\end{eqnarray}
as a functional of the metric and boundary conditions $ \xi^{\mu}=
\xi^{\mu}[\,g, {\rm boundary\; data}]$. The boundary data is an
inalienable
part of the solution of  (\ref{7.6}), and this data is nontrivial
and
nontrivially depends on $\beta$. This means, that iteratively solving
the
equation (\ref{7.6}) we can obtain $ \xi^{\mu}$ as a nonlocal
expansion in
curvatures, but the nontrivial dependence on $\beta$ will enter this
functional
through boundary conditions. Therefore, the dependence of
$\mbox{\boldmath
$\Gamma$}_{\beta}$ on $\beta$ will be induced not only by the metric
argument
of $\mbox{\boldmath $\Gamma$}[\,g\,]$: $\mbox{\boldmath
$\Gamma$}_{\beta}=\mbox{\boldmath $\Gamma$}[\,g^{\beta},\beta\,]$
($\mbox{\boldmath $\Gamma$}_{n}(x_1,...x_n)\equiv\mbox{\boldmath
$\Gamma$}_{n}(\,\beta\,|\,x_1,...x_n)$) and the above mechanism will
break
down, since the first-order term in $(\beta-8\pi M)$ will no longer
be
generated by the Einstein term of the effective action.

Even if this specific mechanism proposed by Susskind does not work,
there  may
be other solutions of the puzzle. But it looks like practically
impossible to
explain the huge entropy of black holes without relating it to the
properties
of vacuum in a strong gravitational field of a black hole and without
identifying the dynamical degrees of freedom of a black holes with
states of
physical fields located inside a black hole.

\section*{Acknolegements}
\hspace{\parindent}
The authors benefitted from helpful discussions with W.Israel ,
E.Martinez and
G.Hayward. The work of A.O.B. was supported by NSERC grants at the
University
of Alberta, while V.P.F. and A.Z. were supported by NSERC under the
Research
Grant OGP0138712.

\appendix
\section{Lichnerowicz \ equation \ and \  the \ geometry \ of
Einstein-Rosen \ bridge}
\setcounter{equation}{0}
\renewcommand{\theequation}{A.\arabic{equation}}
\hspace{\parindent}
Here we show that the three-geometry of a spatial section on which we
define
the no-boundary wavefunction of true physical variables (\ref{3.12}),
$(g^T,\,p_{T})=(h_{ab}^T,\,p_{T}^{ab}$, matter variables) coincides
with the
geometry of the Einstein-Rosen bridge in the lowest order of the
perturbation
theory in $(g^T,\,p_{T})$. This approximation corresponds to a ground
state of
physical excitations (of both matter and gravitational fields) on the
spatial
section with the topology (\ref{3.4}).

Consider three-geometry ${}^3\!g_{ab}$ and define
\begin{equation}\label{A.0}
\tilde{\beta}^{ab}=\epsilon^{aef}\nabla_e [\sqrt{{}^3\!{g} } ({}^3
R_f^b
-\frac{1}{4}\delta^b_f \ {}^3\!R)] .
\end{equation}
York  \cite{York}  showed that $\tilde{\beta}^{ab}$ gives a pure
spin-two
representation of intrinsic geometry. Conditions
$\tilde{\beta}^{ab}=0$
together with $p_{ab}=0$  specify the state where no dynamical
gravitational
perturbations are present. In the absence of matter the  Lichnerowicz
equation
(\ref{3.10}) reduces to the equation
\begin{equation}\label{A.0a}
{}^3\!R =0 .
\end{equation}
Condition $\tilde{\beta}^{ab}=0$ implies that the three-metric is
conformally
flat
\begin{equation}\label{A.0b}
dl^2=\Phi^4 d\tilde l^2_0 =\Phi^4 (dx^2 +dy^2 +dz^2 ) .
\end{equation}
The Lichnerowicz equation (\ref{A.0a}) in this case is equivalent to
the
equation
\begin{equation}\label{A.0c}
\triangle \Phi =0
\end{equation}
for the conformal factor $\Phi$.  A solution which is regular
everywhere is
constant and the corresponding geometry is a flat three-dimensional
space
$R^3$. Non-trivial solutions have singularities. A solution with one
simple
pole generates a three-dimensional space $S^2 \times R^1$ with the
Einstein-Rosen bridge geometry.  We choose coordinates so that the
pole is
located at the origin of coordinates, then we have
\begin{equation}\label{A.0d}
\Phi =1+\frac{M}{2\rho} ,
\end{equation}
where $\rho^2 =x^2 +y^2 +z^2 $.  For this conformal factor the metric
$dl^2$
can be written as
\begin{equation}\label{A.0e}
dl^2=\frac{dr^2}{1-2M/r}+r^2 d\Omega^2 ,
\end{equation}
where $ r\equiv\rho\left( 1+\frac{M}{2\rho} \right) ^2$. A point
$\rho=\infty$
corresponds to spatial infinity of $\Sigma_+$, while a point
$\rho=0$
corresponds to spatial infinity of $\Sigma_-$, the constant $M$ being
the mass
($M_+ =M_- =M$). The important property of the obtained solution
describing the
state without excitation is that the corresponding three-metric is
spherically-symmetric.

The metric (\ref{A.0b}) with (\ref{A.0d}) can be identically
rewritten in the
form in which both spatial infinities are represented in the
completely
symmetric way. To do this we remind that the flat metric is
conformally related
with a metric on a three-sphere $S^3$, so that we have
\begin{equation}
	dl^2 =\tilde{\Phi}^4_0 (d\chi^2 +
	\sin^2{\chi}d\Omega^2)=
	\frac{dr^2}{1-2M/r} +r^2d\Omega^2, \,\,\, ,     \label{A.7}
	\end{equation}
where $\tilde{\Phi}_0$ is a solution of the conformal invariant
equation on the
three-sphere
	\begin{equation}
	(\tilde \Delta-\frac18\,^3\! \tilde R)\, \tilde{\Phi }_0=0,
\label{A.3}
	\end{equation}
which is of the form $\tilde{\Phi}_0=\phi_0  /\sin \chi$, with $\phi
_0
=M^{1/2} [\sin{(\chi /2)}+\cos{(\chi /2)} ]$ .

In the presence of gravitational perturbations and matter  the
Lichnerowicz
equation  (\ref{3.10})  reads
	\begin{equation}
	(\tilde\Delta-\frac18\,^3\!\tilde R)\,
	\tilde{\Phi} =J ,
  \label{A.1}
	\end{equation}
where the source $J$ in terms of conformally transformed variables
looks as
	\begin{equation}
	J=-\frac18\, (^3\tilde g)^{-1}\,
	\ ^3\tilde p^{ab}\,^3\tilde p_{ab}\Phi^{-7}-2\pi\,
	\tilde T^*_*\,\Phi^{-3}  .
\label{A.2}
	\end{equation}
Denote by $G(\mbox{\boldmath $x,x$}')$ the Green function defined as
the
solution of the equation
      \begin{equation}\label{A.2a}
      (\tilde \Delta-\frac18\,^3\! \tilde R)\,G(\mbox{\boldmath
$x,x$}')=-{}^3\!\delta(\mbox{\boldmath $x,x$}')\ .
      \end{equation}
The solution of the equation (\ref{A.1}) can be presented in the form
        \begin{equation}\label{A.2b}
         \tilde{\Phi} =\tilde{\Phi} _0 +\int G(\mbox{\boldmath
$x,x$}' )
        J(\mbox{\boldmath $x$}' ) \tilde{g}^{1 \over 2}
d\mbox{\boldmath $x$}'
{}.
\end{equation}
The solution $\tilde{\Phi} _0$ is invariant with respect to the
reflection
$\chi\rightarrow\pi -\chi$. In general case  $J$ does not obeyes this
property
and the solution $\tilde{\Phi}$ is not invariant under reflection
and
asymptotic values of $M_+$ and $M_-$ are different. In order to
illustrate this
general property we consider here a simple case when $J$ is
spherically
symmetric.

We write $\tilde{\Phi}$ in the form $\tilde{\Phi}=\phi /\sin \chi$ .
The
function $\phi$ obeys the equation
	\begin{equation}
	{d^{2}{\phi} \over d\chi^2}+{1 \over 4}\phi =j\equiv J\sin
\chi . \label{B.9}
	\end{equation}
and has a general solution in terms of  the Green function $G(\chi
,\chi ' )$:
	\begin{eqnarray}
	&&\!\!\!\!\!\!\!\!\!\!\!\!\!\!\!\!\phi (\chi )= \phi_0 (\chi
) +
	\int_0^{\pi} G(\chi ,\chi ' ) j(\chi ' )d\chi '
,\label{B.10}\\
	&&\!\!\!\!\!\!\!\!\!\!\!\!\!\!\!\!G(\chi ,\chi ' )=
	-2\left\{ \theta(\chi -\chi ' )\,\sin (\chi /2) \cos(\chi
'/2) +
	\theta(\chi' -\chi )\,\sin (\chi  '/2) \cos(\chi /2)
\right\}. \label{B.11}
	\end{eqnarray}
The asymptotic masses at  two spatial infinities are
	\begin{equation}
	M_+ =\left.\phi\,{d{\phi} \over d\chi}\, \right|_{\chi=0},
	\hspace{1cm}M_- =
	\left. -\phi\,{d{\phi} \over d\chi} \,\right|_{\chi=\pi} .
\label{A.8}
	\end{equation}
{}From this expression it follows that
	\begin{eqnarray}
	&&\phi (0)=M^{1/2} ,\;\;\;\phi (\pi )=M^{1/2} \;\;\;
	\phi ' (0)=M^{1/2}/2 -\alpha,\;\;\;\phi ' (\pi )=-M^{1/2}/2
+\beta  ,\nonumber
\\
	&&\alpha \equiv\int_0^{\pi} \cos (\chi ' /2)j(\chi ' )d\chi '
,
	\;\;\;\;\beta \equiv\int_0^{\pi} \sin (\chi ' /2)j(\chi '
)d\chi ',
	\end{eqnarray}
whence
	\begin{equation}\label{b.13}
	M_+ -M_-  =2 M^{1/2}(\beta -\alpha ).
	\end{equation}
This relation shows that in the general case the asymmetric
distribution of
matter on the Einstein-Rosen bridge results in different masses $M_+$
and $
M_-$ at two asymptotic infinities. For a known distribution and fixed
$M_+$
the value of   $ M_-$ can be obtained by solving of the  Lichnerowicz
equation.

\section{R-Modes}
\setcounter{equation}{0}
\renewcommand{\theequation}{B.\arabic{equation}}

In this appendix we construct the basis of positive frequency
solutions
\begin{equation}
w_{\lambda}=\frac{1}{\sqrt{4\pi \omega}}\exp (-i\omega t) R_{\omega
lm}(r,\vartheta ,\phi )
\end{equation}
for the scalar field  in the exterior region $R_+$ of the eternal
black hole,
for which spatial functions $R_{\omega lm}$ are real ({\em R-modes}).

By using the separation of variables for the equation $\Box \varphi
=0$ we
write
\begin{equation}\label{a.1}
R_{\omega lm}(r,\vartheta ,\phi )=R_{\omega
l}(r)\hat{Y}_{lm}(\vartheta ,\phi )
,
\end{equation}
where
\begin{equation}\label{a.2}
\hat{Y}_{lm}(\vartheta ,\phi ) =P^m_l  (\vartheta)\left\{
\begin{array}{ll}
\frac{1}{\sqrt{2\pi}} ,&m=0 ,\\
\frac{1}{\sqrt{\pi}}\cos m\phi ,&0< m\le l  ;\\
\frac{1}{\sqrt{\pi}}\sin m\phi  ,&-l\le m<0 .
\end{array}
\right.
\end{equation}
We choose the spherical harmonics $\hat{Y}_{lm}$  to be real so that
the
R-basis will be constructed if  solutions $R_{\omega l}(r)$ of the
radial
equation  (\ref{5.29}) are chosen to be real.   Denote
$\hat{R}_{\omega l}(r)=r
R_{\omega l}(r)$ then the radial equation reads
\begin{equation}\label{a.3}
\frac{d^2 \hat{R}_{\omega l}}{d{r^*}^2}+(\omega^2-V_l
)\hat{R}_{\omega l}=0 \ ,
\label{B4}
\end{equation}
where $r^*=r-2M +2M\ln{[(r-2M)/2M]}$, and
\begin{equation}
V_l =\left( 1-\frac{2M}{r} \right) \left(
\frac{l(l+1)}{r^2}+\frac{2M}{r^3}
\right) .
\end{equation}
For any two solutions of (\ref{B4}) the Wronskian $W[f_1,f_2] \equiv
f_1 ({df_2
\over dr^*}) - f_2 ({df_1 \over dr^* }) =$const .

Functions $\hat{R}_{\omega l}$ have the asymptotics $\exp{(\pm
i\omega r)}$ at
$r\rightarrow\infty$ and $\exp{(\pm i\omega r^*)}$  at $r^*
\rightarrow
-\infty$. We begin by defining so called $UP$-modes which are
specified (for
$\omega >0$) by the asymptotics
\begin{equation}\label{a.4}
\hat{R}_{\omega l}^{up}(r)=\left\{ \begin{array}{ll}
e^{ i\omega  r^*}+r_{\omega l}e^{ -i\omega  r^*},  r^* \rightarrow
-\infty,\\
t_{\omega l}e^{ i\omega  r},  r\rightarrow  \infty .
\end{array}
\right.
\end{equation}
By comparing the Wronskians at  $r^* =\pm \infty$ for
$\hat{R}_{\omega l}^{up}$
and its complex conjugated one gets the standard relations between
reflection
and absorption coefficients
\begin{equation}\label{a.5}
|r_{\omega l}|^2 +|t_{\omega l}|^2 =1 .
\end{equation}

The coefficients of the radial equation are real. That is why
$\hat{R}_{\omega
l}^{down}(r)\equiv \bar{\hat{R}}_{\omega l}^{up}(r)$ is again a
solution. One
has
\begin{equation}\label{a.6}
\bar{\hat{R}}_{\omega l}^{up}(r)=\hat{R}_{-\omega l}^{up}(r) ,
\end{equation}
so that $\bar{r}_{\omega l}=r_{-\omega l}$ and  $\bar{t}_{\omega
l}=t_{-\omega
l}$.  The $Re$- and $Im$-parts of  $\hat{R}_{-\omega l}^{up}(r) $
(for $\omega
>0$) can be used as real basic solutions. The problem is that the
corresponding
solutions $w_{\lambda}$ do not possess the proper normalization
conditions.
Namely one has
\begin{eqnarray}\label{a.7}
(w^{up}_{\omega lm},w^{up}_{\omega 'l'm'})&=&\delta (\omega -\omega
')\delta_{ll'}\delta_{mm'} ,\\
(w^{down}_{\omega lm},w^{down}_{\omega 'l'm'})&=&\delta (\omega
-\omega
')\delta_{ll'}\delta_{mm'} ,\\
(w^{down}_{\omega lm},w^{up}_{\omega 'l'm'})&=&r_{\omega l}\delta
(\omega
-\omega ')\delta_{ll'}\delta_{mm'} .
\end{eqnarray}
Here
\begin{equation}\label{a.8}
(f_1 ,f_2 )=-i\int{(\bar{f}_1 f_{2 ,\mu} -\bar{f}_2 f_{1
,\mu})d\sigma^{\mu}}
\end{equation}
is a scalar product in the space of solutions.

The proper normalization conditions can be satisfied by the following
linear
transformation of the basic functions
\begin{eqnarray}\label{a.9}
\hat{R}_{\omega l}^{up '}&=&a_{\omega l}\hat{R}_{\omega
l}^{up}+b_{\omega
l}\hat{R}_{\omega l}^{down},\\
\hat{R}_{\omega l}^{down '}&=&\bar{b}_{\omega l}\hat{R}_{\omega
l}^{up}+a_{\omega l}\hat{R}_{\omega l}^{down},
\end{eqnarray}
where
\begin{eqnarray}\label{a.10}
a_{\omega l}=\frac{\sqrt{1+|t_{\omega l}|}}{\sqrt{2} |t_{\omega l}|}
,\hspace{1cm}
b_{\omega l}=-\frac{r_{\omega l}}{\sqrt{2} |t_{\omega
l}|\sqrt{1+|t_{\omega
l}|}}.
\end{eqnarray}

The following functions are real and for $\omega >0$ form a proper
normalized
basis
\begin{eqnarray}\label{a.11}
\hat{R}_{\omega l1}^{real}&=&\frac{1}{\sqrt{2}} (\hat{R}_{\omega
l}^{up'}+\hat{R}_{\omega l}^{down'}),\\
\hat{R}_{\omega l2}^{real}&=&\frac{1}{i\sqrt{2}} (\hat{R}_{\omega
l}^{up'}-\hat{R}_{\omega l}^{down'}) .
\end{eqnarray}
To summarize we construct  the basis $\{ w_{\lambda}\}$ \  ($\omega
>0$,
$A=1,2$)
\begin{equation}\label{a.12}
 w_{\lambda}=\frac{1}{\sqrt{4\pi \omega}}\exp (-i\omega t)
R^{real}_{\omega lm
A}(r,\vartheta ,\phi ) ,
\end{equation}
where $R^{real}_{\omega lm A}\equiv  r^{-1} \hat{R}_{\omega
lA}^{real}\hat{Y}_{lm}(\vartheta ,\phi ) $ are real functions.


\begin{thebibliography}{99}
\bibitem{fn1} The analogous divergent expression  arose ealier in the
framework
of "brick wall model"  proposed by 't Hooft (see, also [26]).
Recently Susskind
and Uglum [27] also obtained the analogous divergency and proposed to
relate it
with the renormalization of the gravitational constant. They also
show that in
the superstring theory one can get the final value instead of the
divergency.
\bibitem{fn3} In the case of a rotating black hole the inequality
$E<0$ must be
changed by $E<0, E-L\Omega_H <0$, where $L$ is an angular momentum of
a
particle and $\Omega_H$ is the azimuthal angular velocity of a black
hole. More
accurate general definition of states corresponding to internal
degrees of
freedom of a black hole, which can be used for rotating and charged
black holes
will be given later in the paper.
\end{thebibliography}
\end{document}